\documentclass[prb,twocolumn,floatfix]{revtex4}
\usepackage{amsmath}  % needed for \tfrac, \bmatrix, etc.
\usepackage{amsfonts} % needed for bold Greek, Fraktur, and blackboard bold
\usepackage{graphicx} % needed for figures

\usepackage{bm}

\usepackage{mathptmx, courier, pifont}
\usepackage[scaled=0.92]{helvet}
\usepackage[T1]{fontenc}
\usepackage{textcomp}

\catcode`\@=11
\renewcommand{\ps@myheadings}{\let\@mkboth\@gobbletwo
\def\@oddhead{\hbox{}\hfil {\normalsize\sl\rightmark}\hfil 
{\bf\arabic{page}}\hbox{}}
\def\@oddfoot{}\def\@evenhead{{\bf\arabic{page}}\hfil  %%--> Change is here  
{\normalsize\sl\leftmark} \hfil\hbox                  
{}}\def\@evenfoot{}\def\sectionmark##1{}\def\subsectionmark##1{}}

\newcommand{\yes}{yes}
     
\newcommand{\labelshow}{no}        %%-->  Set default to not show labels

\newcommand{\labelformat}[1]              %%-->  Format for writing labels
           {\{#1\}}     %% {\small [\thin label=#1\thin]$\rightarrow$}

\newcommand{\eq}[1]{%                             
\ifx\labelshow\yes%                                % If \showlabels 
Eq.\ {\{#1\}}(\ref{eq:#1})%
\else%                                         
Eq.\ (\ref{eq:#1})%                % If \noshowlabels(default)
\fi}

\newcommand{\eqnoeq}[1]{%                             
\ifx\labelshow\yes%                                % If \showlabels 
{\{#1\}}(\ref{eq:#1})%
\else%                                         
(\ref{eq:#1})%                % If \noshowlabels(default)
\fi}

\newcommand{\rangeref}[2]{%                             
\ifx\labelshow\yes%                             % If \showlabels 
Eqns.\ ({\{#1\}}%
\ref{eq:#1})--({\{#2\}}% 
\ref{eq:#2})%                         %%--> for range 
\else%                                         
Eqns.\ (\ref{eq:#1})--(\ref{eq:#2})\nobreak   % If \noshowlabels(default)%
\fi}

\newcommand{\reference}[1]{%                             
\ifx\labelshow\yes%                             % If \showlabels 
{\{#1\}}\ \cite{#1}%
\else%                                         
\cite{#1}%              % If \noshowlabels(default)
\fi}

\newcommand{\figgy}[4]{%                             
\ifx\labelshow\yes%                             % If \showlabels 
\begin{figure}[tbp]%      %%-->  #3 is for postscript file 
\vspace{#2}%
\begin{center}\parbox{5 true in}{%
\caption[]{\{#1\} \protect\label{fg:#1} #4}%
}\end{center}%
\end{figure}%
\else%                                       % If \noshowlabels(default)   
\begin{figure}[tbp]%      %%-->  #3 is for postscript file 
\vspace{#2}%
\begin{center}\parbox{5 true in}{%
\caption[]{\small \protect\label{fg:#1} #4}%
}\end{center}%
\end{figure}%
\fi}

\newcommand{\twofiggy}[8]{%                             
\ifx\labelshow\yes%                             % If \showlabels 
\begin{figure}[tbp]%      %%-->  #3 is for postscript file 
\vspace{#2}%
\begin{center}\parbox{5 true in}{%
\caption[]{\small \{#1\} \protect\label{fg:#1} #4}%
}\end{center}%
\vspace{#6}%
\begin{center}\parbox{5 true in}{%
\caption[]{ \small \{#5\} \protect\label{fg:#5} #8}%
}\end{center}%
\end{figure}%
\else%                                       % If \noshowlabels(default)   
\begin{figure}[tbp]%      %%-->  #3 is for postscript file 
\vspace{#2}%
\begin{center}\parbox{5 true in}{%
\caption[]{ \small \protect\label{fg:#1} #4}%
}\end{center}%
\vspace{#6}%
\begin{center}\parbox{5 true in}{%
\caption[]{\small \protect\label{fg:#5} #8}%
}\end{center}%
\end{figure}%
\fi}

\def\picture #1 by #2 (#3){ 
  \vbox to #2{
    \hrule width #1 height 0pt depth 0pt
    \vfill
  \special{picture #3} } }

\def\centerpicture #1 by #2 (#3 scaled #4){
  \dimen0=#1 \dimen1=#2 
  \divide\dimen0 by 1000 \multiply\dimen0 by #4
  \divide\dimen1 by 1000 \multiply\dimen1 by #4
	\ashift{\dimen0}
	\picture \dimen0 by \dimen1 (#3 scaled #4)}

\newlength{\alen}
\newcommand{\ashift}[1]{
\setlength{\alen}{\textwidth}
\addtolength{\alen}{-1#1}
\addtolength{\alen}{-0.5\alen}
\addtolength{\alen}{-8 pt} \hspace{\alen}}

\newcommand{\alignc}{c}
\newcommand{\isok}{c}

\newcommand{\fgcaption}[3]{\renewcommand{\isok}{#1}%
 \ifx  \alignc \isok
\begin{center}%
\parbox{4 in}{\begin{center}%
{\small{\bf\noindent Fig.\  \ref{fg:#2}\hspace{1em}} #3}  \end{center}}%
\end{center}%
\else%
\begin{center}%
\parbox{4 in}{%
{\small{\bf \noindent Fig.\  \ref{fg:#2}\hspace{0.5em}} #3}}%
\end{center}%
\fi}

\newcommand{\fig}[1]{%                             
\ifx\labelshow\yes%                               % If \showlabels 
Fig.~{ \{#1\}} \ref{fg:#1}%
\else%                                         
Fig.~\ref{fg:#1}%                    % If \noshowlabels(default)
\fi}

\newcommand{\figs}[2]{%                             
\ifx\labelshow\yes%                               % If \showlabels 
Figs.~{ \{#1\}} \ref{fg:#1}--{ \{#2\}} \ref{fg:#2}%
\else%                                         
Figs.~\ref{fg:#1}--\ref{fg:#2}%                    % If \noshowlabels(default)
\fi}

\newcommand{\tableinsert}[5]{%     %%--> #1 is label; #2 is size; #3 is for 
\ifx\labelshow\yes%                %%--> postscript file; #4 is upper 
\begin{table}%                     % If \showlabels 
\begin{center}\parbox{5 true in}{%
\caption[]{\{#1\} \protect\label{tb:#1}#4}%
}\end{center}%
\par\vspace {#2}%
\noindent{\footnotesize #5}%
\end{table}%
\else%                                         
\begin{table}%              % If \noshowlabels(default)
\begin{center}\parbox{5 true in}{%
\caption[]{\protect\label{tb:#1}#4}%
}\end{center}%
\par\vspace {#2}%
\noindent{\small #5}%
\end{table}%
\fi}

\newcommand{\tableref}[1]{%                             
\ifx\labelshow\yes%                                % If \showlabels 
Table~{\{#1\}} \ref{tb:#1}%
\else%                                         
Table~\ref{tb:#1}%                   % If \noshowlabels(default)
\fi}%

\newcommand{\heading}[2]{%                             
\ifx\labelshow\yes%                             % If \showlabels 
\section{\{#1\} \protect\label{h:#1}#2}%
\else%                                         
\section{\protect\label{h:#1}#2}%
\fi}%

\newcommand{\subheading}[2]{%                             
\ifx\labelshow\yes%                             % If \showlabels 
\subsection{\{#1\} \protect\label{sh:#1}#2}%
\else%                                         
\subsection{\protect\label{sh:#1}#2}%
\fi}%

\newcommand{\subsubheading}[2]{%
\ifx\labelshow\yes%
\subsubsection{\{#1\} \protect\label{ss:#1}#2}%      %% If \showlabels
\else%
\subsubsection{\protect\label{ss:#1}#2}%                 %%  If \noshowlabels
\fi}

\newcommand{\tightdot}{\kern-.2em\cdot\kern-.2em}  
                                              
\newcommand{\tightcross}{\kern-.2em\times\kern-.2em}  
                                              
\newcommand{\units}[1]{\mbox{\ #1}}
\newcommand{\punits}[2]{\units{#1}^{#2}}

\newcommand{\halfthin}{\kern 0.0834em}
\newcommand{\neghalfthin}{\kern -0.0834em}

\newcommand{\quarterthin}{\kern 0.0417em}
\newcommand{\negquarterthin}{\kern - 0.0417em}

\newcommand{\infinity}{\infty}

\newcommand\Script[1]{{\cal #1}}

\newcommand{\onematrix}[2]{\begin{small}\left(\begin{array}{c}#1\\#2\end{array}\right)
\end{small}}
\newcommand{\twomatrix}[4]{\begin{small}\left(\begin{array}{cc}#1&#2\\#3&#4\end{array}
           \right)\end{small}}

\newcommand{\unittwo}{\begin{small}
\left(\begin{array}{cc}1&0\\0&1\end{array}\right)\end{small}}

\newcommand{\isotope}[2]{\mbox{$^{#1}$#2}}

\newcommand{\trace}{\mbox{tr} \thinspace}
\newcommand{\diag}{\mbox{diag} \thinspace}

\newcommand{\muchgreater}{{\ \scriptstyle >\hspace{-0.2em}>\ }}
\newcommand{\muchless}{{\ \scriptstyle <\hspace{-0.2em}<\ }}

\newcommand{\bra}[1]{\langle#1|}
 \newcommand{\ket}[1]{|#1\rangle}

\newcommand{\tsub}[1]{_{\mbox{\scriptsize#1}}}

\newcount\hours
\newcount\minutes
\newcount\tmp
\hours=\time
\divide\hours by 60
\tmp=\hours
\minutes=\time
\multiply\tmp by 60
\advance\minutes by -\tmp
\newcommand{\timenow}{\number\hours:\ifnum\number\minutes>9%
{\number\minutes}\else%
0\number\minutes\fi}

\newcommand{\datefile}{\date{\vspace{10pt}%\normalsize
            \small\sc  
            ---\thinspace Printed from File: {\small
           \lowercase{\jobname}.tex at \timenow}
            \thinspace---\vphantom{\bigg[} 
            \\ %\normalsize
             \small ---\today---}}

\newcommand{\mnote}[1]%     %%-->  To produce margin notes
{\setlength{\marginparwidth}{50pt}%
\setlength{\marginparsep}{10pt}%
\marginpar{\scriptsize\em#1}}

\newcounter{exercisenumber}

\def\pointsize@{10}
\def\fivepoint{\def\pointsize@{5}%
 \normalbaselineskip7\p@
 \abovedisplayskip7\p@ plus1.8\p@ minus5.4\p@
 \belowdisplayskip7\p@ plus1.8\p@ minus5.4\p@
 \abovedisplayshortskip\z@ plus1.8\p@
 \belowdisplayshortskip2.6\p@ plus1.8\p@ minus1.0\p@
 \textfont@\rm\fiverm
 \textfont@\it\fivei
 \textfont@\bf\fivebf
 \ifsyntax@\def\big##1{{\hbox{$\left##1\right.$}}}\else
 \let\big\eightbig@
 \textfont\z@\fiverm \scriptfont\z@\fiverm \scriptscriptfont\z@\fiverm
 \textfont\@ne\fivei \scriptfont\@ne\fivei \scriptscriptfont\@ne\fivei
 \textfont\tw@\fivesy \scriptfont\tw@\fivesy \scriptscriptfont\tw@\fivesy
 \textfont\thr@@\tenex \scriptfont\thr@@\tenex \scriptscriptfont\thr@@\tenex
 \textfont\itfam\fivei
 \textfont\bffam\fivebf \scriptfont\bffam\fivebf \scriptscriptfont\bffam\fivebf
 \fi
 \setbox\strutbox\hbox{\vrule height 5\p@ depth2\p@ width\z@}%
 \setbox\strutbox@\hbox{\vrule height 4\p@ depth1\p@ width\z@}%
 \normalbaselines\fiverm\ex@=.2326ex}

\def\boxit#1{\vbox{\hrule\hbox{\vrule\kern3pt
     \vbox{\kern3pt#1\kern3pt}\kern3pt\vrule}\hrule}}

\def\yspacef{\kern 0.6\pointsize@\p@}
\def\yspaceo{\kern 0.3em }

\def\cstok#1{
\if5\pointsize@ \let\yspace\yspacef \else
 \let\yspace\yspaceo \fi
  \leavevmode\hbox{\kern-.4pt\vrule\vtop{\vbox{\hrule
      \kern 0.1\pointsize@\p@
        \hbox{\vphantom{\rm/}\yspace{#1}\yspace}}
      \kern 0.1\pointsize@\p@\hrule}\vrule}}

\def\Young#1{\null\,\vcenter{\normalbaselines\m@th
     \if8\pointsize@ \baselineskip=8pt \lineskip=-0.4pt
     \else \baselineskip=10pt \lineskip=-0.4pt \fi
     \if5\pointsize@ \baselineskip=5pt \lineskip=-0.4pt\fi
    \ialign{$##$&&$##$\crcr
      \mathstrut\crcr\noalign{\kern-\baselineskip}
      #1\crcr\mathstrut\crcr\noalign{\kern-\baselineskip}}}\,}

\newcommand{\singlefig}[6]{%
\begin{figure}[tpb]\vspace{#3}%
\includegraphics*[scale=#5]{#2}%
\caption{\label{#1} #6}%
\vspace{#4}%
\end{figure}}

\newcommand{\doublefig}[6]{%
\begin{figure*}[tpb] \vspace{#3}%
\includegraphics*[scale=#5]{#2}%
\caption{\label{#1} #6}
\vspace{#4}
\end{figure*}}

\newcommand{\thetav}{\theta}
\newcommand{\thetam}{\theta\tsub m}
\newcommand{\dotthetam}{\dot\theta\tsub m}
\newcommand{\U}{\twomatrix{\cos\thetav}{\sin\thetav}{-\sin\thetav}{\cos\thetav}}
\newcommand{\Udag}{\twomatrix{\cos\thetav}{-\sin\thetav}{\sin\thetav}{
\cos\thetav} }

\renewcommand{\onematrix}[2]{\begin{pmatrix}#1\\#2\end{pmatrix}}
\renewcommand{\twomatrix}[4]{\begin{pmatrix}#1&#2\\#3&#4\end{pmatrix}}

\renewcommand{\eq}[1]{Eq.~(\ref{#1})}
\renewcommand{\fig}[1]{Fig.~\ref{#1}}

\newcommand{\dalem}{{\Box}}

\newcommand{\sprod}[2]{\bm#1\cdot\bm#2}

\newcommand{\diffelement}[1]{d#1}

\newcommand{\resonanceDensity}{\mbox {$n_{\rm e}^{\rm\scriptscriptstyle R}$}}

\newcommand{\pee}[1]{P(\nu\tsub e{\,\scriptstyle \rightarrow}\,\nu\tsub e,\,#1)}
\newcommand{\pemu}[1]{P(\nu\tsub e{\,\scriptstyle\rightarrow}\,\nu_\mu,\,#1)}

\newcommand{\peec}{\bar P(\nu\tsub e {\,\scriptstyle\rightarrow\,}\nu\tsub e)}
\newcommand{\pemuc}{\bar P(\nu\tsub e {\,\scriptstyle\rightarrow\,}\nu_\mu)}

\newcommand{\mattermass}[1]{\ket{\nu_{\scriptscriptstyle 
#1}^{\scriptscriptstyle\rm m}}}

\newcommand{\cerenkov}{Cherenkov}

\usepackage{todonotes}
\usepackage{color}
\definecolor{red}{rgb}{1,0.,0}

\begin{document}

\title{A Basic Introduction to the Physics of Solar Neutrinos}

\author{Mike Guidry}
\email{guidry@utk.edu} 
\affiliation{Department of Physics and Astronomy, University of Tennessee, 
Knoxville, TN 37996--1200}

\author{Jay Billings}
\email{billingsjj@ornl.gov} 
\affiliation{Oak Ridge National Laboratory, Oak Ridge, Tennessee 37830}

\date{\today}

\begin{abstract}

A comprehensive introduction to the theory of the solar neutrino problem  is 
given that is aimed at instructors  who are not experts in quantum field theory 
but would like to incorporate these ideas into instruction of advanced 
undergraduate or beginning graduate students in courses like astrophysics or 
quantum mechanics; it is also aimed at the inquisitive  student who would like 
to learn this topic on their own.  The presentation assumes as theoretical 
preparation only that the reader is familiar with the basics of quantum 
mechanics in Dirac notation and elementary differential equations and matrices.  

\end{abstract}

\maketitle

%\tableofcontents

\section{Introduction} 

The resolution of the solar neutrino problem in terms of neutrino flavor 
oscillations has had an enormous impact on both astrophysics and elementary 
particle physics. On the one hand, it has demonstrated that the {\em Standard 
Solar Model (SSM)} works remarkably well, and has suggested profitable new 
directions in a number of astrophysics subfields, while on the other hand it has 
demonstrated conclusively that there is physics beyond the {\em Standard Model  
(SM)} of elementary particle physics and has suggested possible clues for what 
that new physics could be.  It is fitting then that the importance of these 
ideas was recognized in both the 2002 and 2015 Nobel Prizes in 
Physics.  Presumably for these and related reasons, we have found that solar 
neutrino oscillations attract more enthusiastic responses from students in 
advanced undergraduate and graduate courses on stellar structure and stellar evolution than 
any other topic.  

Neutrino oscillations in vacuum and in matter are described extensively in the 
journal literature and in textbooks but most of those discussions are either 
conceptual overviews, \cite{walt2004,hax00,hax04} necessarily omitting a systematic 
mathematical formulation of solar neutrino theory, or treat solar neutrino theory 
extensively but are written assuming that the reader has a substantial 
background in quantum field theory of the weak interactions. 
\cite{kuo1989,smir2003,blen2013} Most advanced undergraduate and beginning graduate students interested in this 
problem, and instructors who are not experts but wish to understand and 
incorporate this topic into their courses, will find neither to be ideal 
sources by themselves. The first introduces the 
concepts well but omits much of the deep mathematical `why' and `how' of these beautiful 
ideas; the second formulates the theory more systematically but typically 
requires a substantial remedial effort for the intelligent non-expert to come up 
to the required speed, often omitting important steps in derivations that are 
(relatively) obvious to the cognoscenti, but perhaps not to others.

This paper attempts to provide a systematic introduction to solar neutrino 
oscillations and the resolution of the solar neutrino problem at a level that 
will satisfy non-specialist instructor and student alike within an accessible 
mathematical framework.  It assumes only that the reader is familiar initially 
with mathematics and basic quantum mechanics at a level characteristic of an 
advanced undergraduate physics major.  No particular background in solar 
astrophysics, elementary particle physics, or relativistic quantum field theory is assumed, as the required concepts 
are introduced as part of the presentation.

\section{\label{neutrinosAndSM} Solar Neutrinos and the Standard Model}

We begin with a concise overview of solar neutrinos and of the Standard 
Model of elementary particles, and of why neutrino observations led to a 
``solar neutrino problem'' implying that either our understanding of the Sun, 
or our understanding of the neutrino, or perhaps both, required fixing.  The 
approach will be decidedly pragmatic, describing only  aspects of these two 
fields that are directly relevant to the task at hand.

\subsection{\label{solarNeutrinos} Neutrinos in the Standard Solar Model}

The Sun is powered by nuclear reactions taking place in its core that convert 
hydrogen into helium and release energy in the process.  Two different 
mechanisms can accomplish this:  (1)~the {\em PP chains,} and (2)~the {\em CNO 
cycle.}  At its present core temperature almost 99\% of the Sun's energy is 
provided by the PP chains, so they will be our focus. The details of 
these nuclear reactions are well documented and will not be of direct concern 
here, except to note that some of these reactions are weak interactions that 
produce neutrinos and that these can be detected on Earth. This is of large 
importance in astrophysics because the neutrinos are produced in the core and 
leave the Sun at essentially the speed of light, with little probability to 
react with the solar matter.  Thus their detection on Earth provides a snapshot 
of conditions in the solar core approximately 8.5 minutes prior to detection, 
and a stringent test of solar models specifically and, by implication, of more 
general models for  stellar structure and stellar evolution.  

\begin{figure}[t]
\centering
\includegraphics[scale=0.63]{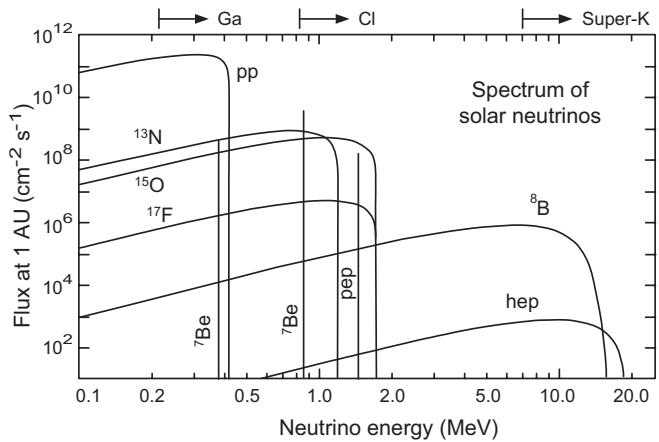}
\caption{The SSM neutrino spectrum, with fluxes scaled to the distance of Earth 
from the Sun. \cite{bah01}  Labels refer to specific reactions in the PP chains 
and CNO cycle. The sensitive regions for various experiments are indicated above 
the graph, with  Ga denoting the GALLEX and SAGE gallium detectors,  Cl denoting 
the Homestake (Davis) chlorine detector,  and Super K denoting the Super 
Kamiokande water \cerenkov\ detector (see Table \ref{snusExp}).} 
\label{solarNeutrinoSpectrum}
\end{figure}

\begin{figure}[t]
\centering
\includegraphics[scale=0.74]{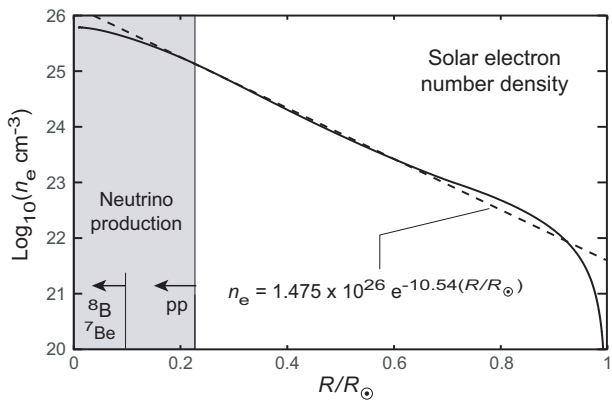}
\caption{Electron number density as a function of fractional solar 
radius from the Standard Solar Model. \cite{bah01} The 
dashed line is an exponential approximation that will be employed in discussing 
the MSW effect. Regions of primary neutrino production in the PP chains are 
indicated, using labels from \fig{solarNeutrinoSpectrum}.}
\label{electronDensitySSM}
\end{figure}

For our specific purposes all that is relevant concerning these solar
neutrino-producing reactions are the flavor (neutrino type) and spectrum of the 
neutrinos that are produced, and the electron density distribution in the solar 
interior.  These may be obtained from the semi-phenomenological {\em Standard 
Solar Model,} which combines the theory of stellar structure and evolution 
with the values of key parameters inferred from measurements or theory to 
provide a comprehensive and predictive model for the structure and evolution of 
the Sun. \cite{bah89} The spectrum of solar neutrinos predicted by the SSM is 
displayed in Fig.~\ref{solarNeutrinoSpectrum},
and the electron number density as a function of 
radius from the SSM is displayed in  Fig.~\ref{electronDensitySSM}.

When solar neutrinos are detected on Earth by chemistry-based or normal-water 
\cerenkov\ detectors, the flux is found to be systematically less than that 
predicted by the SSM  (see Fig.~\ref{solarNeutrinoSpectrum}), in an amount that 
depends on the energy of the neutrinos; Table \ref{snusExp} illustrates.%
\begin{table*}[t]
\centering
\caption{Solar neutrino fluxes from various experiments compared with a Standard 
Solar Model (SSM) calculated in Ref.~\onlinecite{bah95}.  All fluxes are in 
solar neutrino units (SNU), except the result from 
Super Kamiokande. (1 SNU is the neutrino flux that would produce 
$10^{-36}$ interactions per target atom per second.)  Experimental 
uncertainties include systematic and statistical contributions. 
%\fixme{How does 1 SNU relate to $10^6\units{cm$^{-2}$s$^{-1}$}$?}
}
\begin{ruledtabular}
\begin{tabular}{llll}
Experiment & Observed flux  &  SSM  & Observed/SSM
\\
\hline
Homestake & $2.54\pm0.14\pm0.14 \units{SNU}$ & $9.3^{\ +1.2}_{\ -1.4}$ 
          & $0.273 \pm 0.021$
\\
SAGE & $72^{\ +12 \ +5}_{\ -10\  -7} \units{SNU}$ & $137^{\ +8}_{\ -7}$ 
          & $0.526 \pm 0.089$
\\
GALLEX & $69.7 \pm 6.7 ^{\ +3.9}_{\ -4.5} \units{SNU}$ & $137^{\ +8}_{\ -7}$ 
          & $0.509 \pm 0.089$
\\
Super Kamiokande & $2.51^{\ +0.14}_{\ -0.13} \ (10^6\units{cm$^{-2}$s$^{-1}$})$ 
          & $6.62^{\ +0.93}_{\ -1.12}$ 
         & $0.379\pm 0.034$
\\
\end{tabular}
\end{ruledtabular}
\label{snusExp}
\end{table*}
Typically for the highest neutrino energies the observed flux is around $30\%$ of that 
expected while for the lowest energies about $50\%$ the expected number are 
seen.  This is the famous {\em solar neutrino problem.}  The reproducibility in 
independent experiments of the measured flux suppression from that expected 
indicates that the issue is real, and implies some combination of (1)~the 
Standard Solar Model fails to describe properly the conditions in the Sun's core 
(we don't understand the Sun), or (2)~the properties of neutrinos differ from 
that assumed in the prediction of Fig.~\ref{solarNeutrinoSpectrum} (we don't 
understand the neutrino). To investigate these possibilities, let us now 
consider the description of neutrinos and the weak interactions in the Standard 
Model of elementary particle physics that is implicit in the prediction of 
Fig.~\ref{solarNeutrinoSpectrum}.

\subsection{\label{SM} The Standard Model and Weak Interactions}

The Standard Model of elementary particle physics is the most comprehensive and 
well-tested theory yet conceived in theoretical physics, and it accounts for an 
extremely broad range of phenomena.  However, our purposes are specific and 
require only four basic ingredients and one speculation from the SM:  (1)~the 
generational structure of its elementary fermions (matter fields) and associated lepton-number 
conservation laws (see \S\ref{generationalStructure}), (2)~the corresponding implication that neutrinos must be 
identically massless and that only left-handed neutrinos and right-handed 
antineutrinos enter the weak interactions (which builds the maximal breaking of 
parity symmetry into the weak interactions by fiat), (3)~the empirical 
observation that in the quark sector the weak eigenstates are not congruent with 
the mass eigenstates, implying {\em flavor mixing} (see \S\ref{quarkFlavorMixing} and \S\ref{leptonicFlavorMixing}), (4)~the 
long-standing speculation \cite{bono2005} that flavor oscillations might occur 
in the neutrino sector also (which would imply finite masses for neutrinos and 
thus a violation of the minimal Standard Model) (see \S\ref{nvacuumOscillations}), and (5)~that the weak 
interactions can occur through {\em charged currents} mediated by the $W^\pm$ 
bosons and {\em neutral currents} mediated by the $Z^0$ boson (see \S\ref{2neutrinoVac}).

\subsubsection{\label{generationalStructure} Generational Structure}

\begin{figure}[t]
\centering
\includegraphics[scale=0.42]{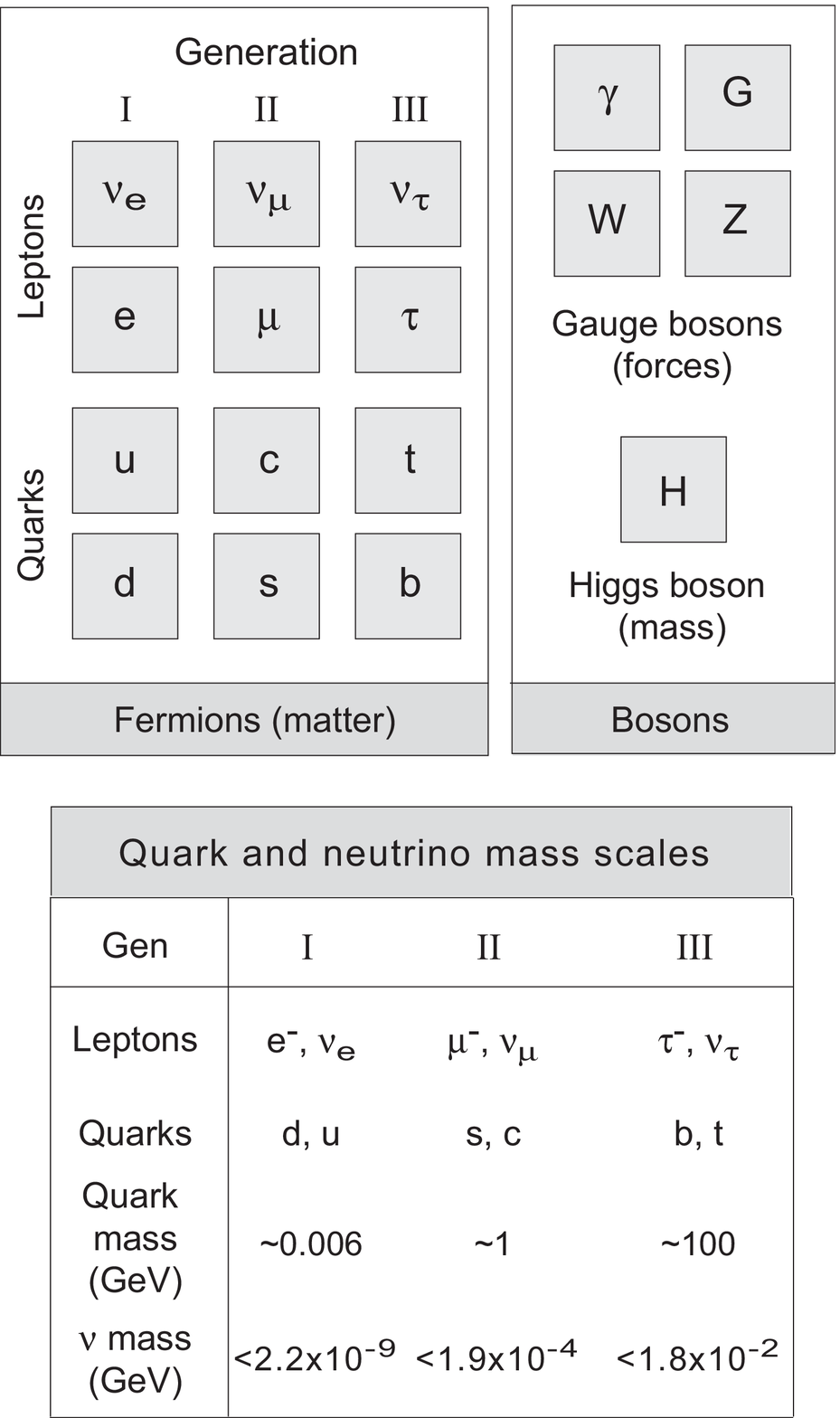}
\caption{Particles of the Standard Model and  characteristic mass scales in the 
quark and neutrino sectors for each  generation. \cite{PDG}}
\label{particleZoo}
\end{figure}

The fermions of the Standard Model are grouped into three 
families or generations, as displayed in Fig.~\ref{particleZoo}.
In the SM, interactions across family lines are forbidden.  For the leptons this 
is implemented formally by assigning a lepton family number to each particle and 
requiring that it be conserved by interactions.
The term {\em flavor} is used to distinguish the quarks and leptons of one 
generation from another.  Thus,  $(\nu\tsub e$, $\nu_\mu,\nu_\tau)$ are 
different flavors of neutrinos, and (u, d, s, \ldots) are
different flavors of quarks.

Also displayed in Fig.~\ref{particleZoo} are characteristic mass scales for 
quarks and neutrinos within each generation.  The neutrino masses are upper 
limits, since no neutrino mass has yet been measured directly. They are seen to 
be tiny on a mass scale set by the quarks of a generation, which is a major 
unresolved issue.  There is no known fundamental reason why the neutrino masses 
should be identically zero (as assumed phenomenologically in the SM), and if 
they are not identically zero, then why are they so small?

\subsubsection{\label{quarkFlavorMixing} Flavor Mixing in the Quark Sector}

The Standard Model  is made richer and more complex  by the  experimental 
observation of {\em flavor mixing} in the quark sector. For quarks the 
mass eigenstates and the weak eigenstates are not equivalent, which means that 
the quark states that enter the weak interactions (flavor eigenstates) are 
generally linear combinations of the mass eigenstates (eigenstates of 
propagating quarks). For example, restricting to the first two 
generations, it is found that the d and s quarks enter the weak interactions in 
the ``rotated'' linear combinations $d\tsub c$ and $s\tsub c$ defined by  the 
matrix equation
\begin{equation}
\begin {pmatrix}
d\tsub c \\ s\tsub c
\end {pmatrix}
=
\begin{pmatrix}
\cos\theta\tsub c & \sin\theta\tsub c
\\
-\sin\theta\tsub c & \cos\theta\tsub c
\end{pmatrix}
\begin{pmatrix}
d \\ s
\end{pmatrix}
\label{cabibboRot}
\end{equation}
where $d$ and $s$ are mass-eigenstate quark fields and the $2\times 2$ unitary 
matrix is parameterized by $\theta\tsub c$, which is termed the {\em mixing angle} or the {\em Cabibbo 
angle}.  Data require the mixing angle to be rather 
small: $\theta\tsub c \sim 13^\circ$.  In the more general case of three 
generations of quarks, weak eigenstates are described by a $3 \times 3$ mixing 
matrix called the {\em Cabibbo--Kobayashi--Maskawa or CKM matrix,} which has 
three real mixing angles and one phase.  There is little fundamental 
understanding of this quark flavor mixing but data clearly require it.

\subsubsection{\label{leptonicFlavorMixing} Flavor Mixing in the Leptonic 
Sector?}

Observation of flavor mixing among quarks raises the question of whether similar 
mixing could occur among the neutrino flavors.  This is irrelevant in the Standard Model 
because to be observable the neutrino flavors undergoing mixing must have different 
masses, which is impossible in the SM where all neutrinos 
necessarily are identically massless. The possibility of flavor oscillations 
among neutrinos is intriguing because it could have large implications both for 
astrophysics and for particle physics.  First, the Sun produces electron neutrinos 
$\nu\tsub e$ and the initial neutrino detection experiments that found the solar 
neutrino problem were sensitive essentially only to $\nu\tsub e$.  If solar neutrinos can 
undergo flavor oscillations, then some $\nu\tsub e$ produced by the Sun could 
reach the Earth as muon neutrinos $\nu_\mu$ or tao neutrinos $\nu_\tau$ that 
would not be detected, thus possibly accounting for the solar neutrino deficit 
exhibited in Table \ref{snusExp}.  Second, observation of neutrino oscillations 
would imply that at least one neutrino flavor has a finite mass, indicating that 
there must be physics beyond the Standard Model. Thus motivated, let us develop 
a theory of flavor oscillations for neutrinos.

\section{\label{nvacuumOscillations} Neutrino Vacuum Oscillations}

There are three known neutrino (and three antineutrino) flavors,  but phenomenology suggests that solar neutrinos 
are well understood in terms of a simple mixing model for two flavors that 
will be employed here. We address first the simplest case of flavor 
oscillations for neutrinos propagating in vacuum. Then  the more difficult (and 
interesting) issue of how the coupling of neutrinos to matter alters flavor 
oscillations will be taken up. In much of this development the convention that 
is standard in particle physics of choosing {\em natural units,} where both 
$\hbar$ and the speed of light $c$ are set to one, will be employed. The Appendix illustrates the relationship between natural units and standard units. 

\subsection{\label{2neutrinoVac} Mixing for Two Neutrino Flavors}

To be definite the two flavors will be assumed to correspond to the electron 
neutrino with wavefunction $\nu\tsub e \equiv \ket{\nu\tsub e}$ and the muon neutrino with wavefunction $\nu_\mu \equiv \ket{\nu_\mu}$, but the formalism could 
be applied to the mixing any two flavors. By analogy with Eq.~(\ref{cabibboRot}) 
for quarks, the flavor eigenstates $\nu\tsub e$ and $\nu_\mu$ are assumed to be 
related to the mass eigenstates $\nu_1 \equiv \ket{\nu_1}$ and $\nu_2 \equiv \ket{\nu_2}$ through the matrix 
transformation
\begin{equation}
\begin{pmatrix}
   \nu\tsub e \\ \nu_\mu
 \end{pmatrix}  
   = U(\thetav) 
   \begin{pmatrix}
   \nu_1 \\ \nu_2
   \end{pmatrix}
   =
   \begin{pmatrix}
   \cos\thetav & \sin\thetav
   \\
   -\sin\thetav & \cos\thetav
   \end{pmatrix}
   \begin{pmatrix}
   \nu_1 \\ \nu_2
   \end{pmatrix}  ,
   \label{2flavorVac}
\end{equation}
where $\thetav$ is a phenomenological vacuum mixing angle chosen 
to lie in the range $0-45^\circ$, and the matrix $U$ is parameterized by the 
single real angle $\thetav$.
The matrix $U$ is unitary (and orthogonal), so $U U^\dagger = U^\dagger U = 
1$, where $U^\dagger$ is the hermitian conjugate of $U$ (in a matrix representation, 
interchange rows and 
columns, and complex conjugate all elements).  This may be used to invert 
Eq.\ (\ref{2flavorVac}) and express the mass eigenstates
 as a 
linear combination of the flavor eigenstates:
\begin{equation}
   \onematrix{\nu_1}{\nu_2} =
U(\thetav)^{\dagger}\onematrix {\nu\tsub e}{\nu_\mu} =
      \twomatrix {\cos \thetav}{-\sin \thetav}
      {\sin \thetav}{\cos\thetav}
      \onematrix {\nu\tsub e}{\nu_\mu}.
   \label{2massVac}
\end{equation}

Assuming that the mass is non-zero for at least one neutrino flavor, the different mass 
eigenstates will have slightly different energies as neutrinos propagate.  Thus, the probability of detecting a specific neutrino 
flavor will oscillate with time, or with  distance traveled.  
From relativistic quantum field theory applied to left-handed neutrinos
(see Section \ref{LHprop}), the mass eigenstates evolve with time $t$ 
according to
\begin{equation}
   \ket{\nu_i(t)} = e^{-iE_it} \ket{\nu_i(0)},
   \label{timeEvolveFlavor}
\end{equation}
where the index $i$ labels mass eigenstates of energy $E_i$, so
the time evolution of the $\nu\tsub e$ state in Eq.~(\ref{2flavorVac}) will
be given  by
\begin{equation}
      \ket{\nu(t)} = \cos \thetav e^{-iE_1t} \ket{\nu_1(0)}
                  + \sin\thetav e^{-iE_2t} \ket{\nu_2(0)},
   \label{vosc2}
\end{equation}
and from the transformation (\ref{2massVac}) this may be expressed as the mixed-flavor state
\begin{align}
      \ket{\nu(t)} &= (\cos^2\thetav e^{-iE_1t}
      + \sin^2\thetav e^{-iE_2t}) \ket{\nu\tsub e}
      \nonumber
      \\
      &\quad+ \sin\thetav \cos\thetav 
      (-e^{-iE_1t} + e^{-iE_2t}) \ket{\nu_\mu}.
   \label{vosc3}
\end{align}
Taking the overlap of this expression 
 with the flavor eigenstates $\ket{\nu\tsub e}$ and $\ket{\nu_\mu}$, the probabilities that a neutrino
initially in an electron neutrino flavor state will remain an electron 
neutrino, or instead
be converted to a muon neutrino after a time $t$, are given respectively by
\begin{subequations}
\begin{align}
   \pee t &= |\bra{\nu\tsub e}\nu(t)\rangle|^2
   \nonumber
   \\
      &= 1 - \tfrac12 \sin^2(2\thetav) [ 1-\cos(E_2 - E_1)t] ,
   \\
   \pemu t
&= |\bra{\nu_\mu}\nu(t)\rangle|^2 
\nonumber
\\[3pt]
      &= \tfrac12 \sin^2 (2\thetav) [ 1-\cos(E_2-E_1)t],
\end{align}
   \label{vosc5}%
\end{subequations}
with the sum of the two probabilities necessarily equal to unity.

\subsection{\label{vacOsc} The Vacuum Oscillation Length}

Equations (\ref{vosc5}) define the essence of vacuum neutrino oscillations in a 2-flavor model.  However, it is normal to put these results into a more practical form through the following considerations.
Neutrinos have at most a tiny mass $m_i$, so  $E_i \muchgreater m_ic^2$ and
the relativistic energy expression may be approximated using a binomial expansion,
\begin{equation}
E_i = (p^2 + m_i^2)^{1/2} \simeq p + \frac{m_i^2}{2p}.
\end{equation}
Thus, assuming that  $E_1\sim E_2 \equiv E$,
\begin{equation}
   E_2 - E_1 \simeq \frac{\Delta m^2}{2E}
\qquad
\Delta m^2 \equiv m_2^2 - m_1^2 .
   \label{vosc6}
\end{equation}
The probability for flavor survival and flavor conversion as a function of 
distance traveled $r \sim ct$ may then be expressed as
\begin{subequations}
\begin{gather}
\pee r
      = 1 - \sin^2 2\thetav \sin^2 \left( \displaystyle\frac{\pi r}{L} \right) ,
\\[3pt]
\pemu r
      = \sin^2 2\thetav  \sin^2 \left( \displaystyle\frac{\pi r}{L} \right) ,
\end{gather}
\label{flavorProbs}%
\end{subequations}
where $\thetav$ is the (vacuum)  mixing angle and the (vacuum) 
{\em oscillation length} $L$, defined by
\begin{equation}
   L \equiv \frac{4\pi E}{\Delta m^2} = \frac{4\pi E \hbar c}{\Delta m^2 c^4}
,
   \label{vosc8}
\end{equation}
is the distance required for one one complete flavor 
oscillation. [In Eq.\ (\ref{vosc8}) $L$ is expressed first in $\hbar = c = 1$ units and then in normal ``engineering units'' with factors of $\hbar$ and $c$ restored; see further discussion in the Appendix.] Neutrino oscillations for a 2-flavor model 
using these formulas are illustrated in Fig.~\ref{vacuumOscillations}.
\singlefig
{vacuumOscillations} 
{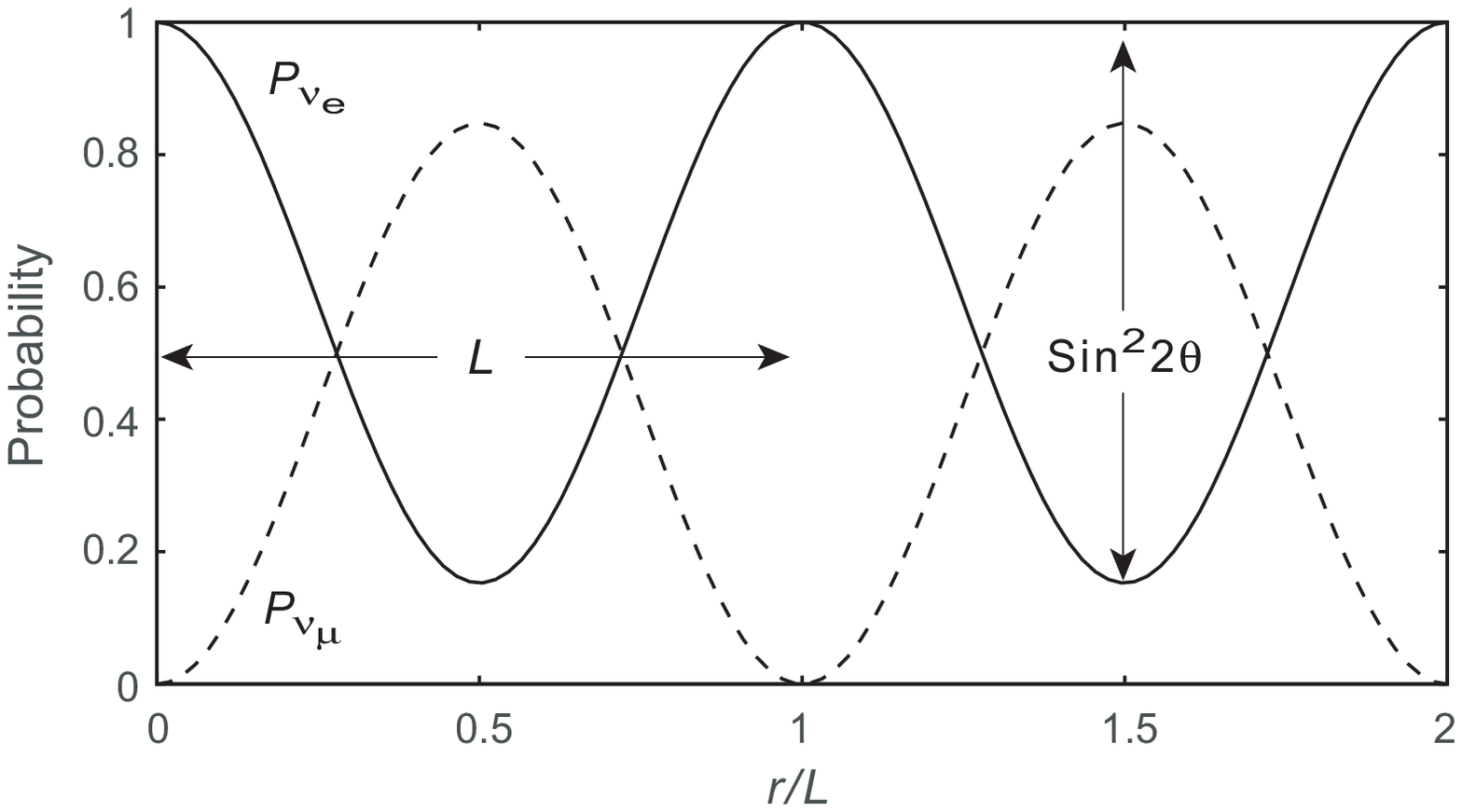}
{0pt}
{0pt}
{0.53}
{Neutrino vacuum oscillations in a 2-flavor model as a function of distance 
traveled $r \sim ct$, with $r$ in units of the vacuum oscillation length $L$. The 
probability as a function of $r$ to be an electron neutrino is denoted by 
$P_{\nu\tsub e}$ and that to be a muon neutrino by $P_{\nu_\mu}$.  The period of 
the oscillation is $L$ and its amplitude is $\sin^22\thetav$, where $\thetav$ is 
the vacuum mixing angle. In this calculation  $\thetav = 33.5^\circ$, the 
neutrino energy is $E = 5 \units{MeV}$, the difference in squared masses for the 
two flavors is $\Delta m^2 c^4 = 7.5 \times 10^{-5} \units{eV}^2$, and the 
corresponding oscillation length $L$ is $165.3 \units{km}$.
}

\subsection{\label{timeAveragedVac} Time-Averaged or Classical Probabilities}

In observations the oscillation length may be smaller than the 
uncertainties in position for emission and detection of neutrinos.  This will 
usually be the case for solar neutrinos, where oscillation lengths are 
typically several hundred kilometers but there are  
variation of thousands of kilometers in the distance traversed between production in the Sun and detection on Earth. If the 
oscillation length is less than the averaging introduced by the preceding 
considerations, the detectors will see a distance (or time) average of 
\eq{flavorProbs}. Denoting the averaged probability by a bar gives 
\begin{equation}
\begin{array}{c}
\peec=
1 - \tfrac12 \sin^2 2\thetav 
\quad
\pemuc=
      \tfrac12 \sin^2 2\thetav,
\end{array}
\label{flavorProbsAVG}
\end{equation}
for the probability (\ref{flavorProbs}) of detecting the two flavors 
distance-averaged over the oscillating factor. This may also be viewed as the 
{\em classical probability,} since the same formula results if the quantum 
interference resulting from squaring the sum of probability amplitudes is removed from the probability.  For two 
flavors the instantaneous probability to remain a $\nu\tsub e$ can approach zero 
if the mixing angle is large (see \fig{vacuumOscillations}) but the {\em 
average} survival probability has a lower limit of $\frac12$ for two flavors.  
The lower limit is  $n^{-1}$ for $n$ flavors, but that limit is
realized only for a flavor mixture tuned to maximal mixing. 

\section{\label{sna1.7} Neutrino Oscillations in Matter}
 
The preceding formalism describes propagation of neutrinos undergoing flavor oscillations in a 2-flavor model in the approximate vacuum from the surface of the Sun to the Earth, but those neutrinos must also propagate through matter from the central regions of the Sun to the surface.
Electron neutrinos couple more strongly to normal matter than do other flavor 
neutrinos because electron neutrinos and the particles making up normal matter 
all reside in the first generation of the Standard Model, but the muon and tau 
neutrinos are in the second and third generations, respectively. It will now be 
shown that this flavor-dependent interaction with the medium alters the 
effective 
mass of a propagating neutrino  differently for electron neutrinos than for 
other flavors, which changes the effective mass-square difference in \eq{vosc6} 
and influences the oscillation non-trivially.

\subsection{\label{sna1.7.2}Propagation of Neutrinos in Matter}

Let us 
consider the effect that interaction with matter can have on  neutrino 
flavor oscillations. \cite{wol78,mik86,bet86,kuo1989,bah89} 
%
%\subsubsection{Matrix Elements for Interaction with Matter}
%
For the low-energy neutrinos found in the Sun inelastic scattering is negligible and electron 
neutrinos in the Sun interact only through {\em elastic forward scattering,} 
mediated by both charged and neutral weak currents. It is instructive to view 
these interactions in terms of Feynman diagrams, which are pictorial 
representations of quantum-mechanical matrix elements.  The Feynman 
diagrams relevant for our discussion are displayed in \fig{MSWFeynman}. For muon or tau neutrinos the 
charged-current diagram is forbidden by lepton family number conservation, so 
only the neutral current can contribute, as illustrated in \fig{MSWFeynman}(b).
\singlefig
{MSWFeynman} 
{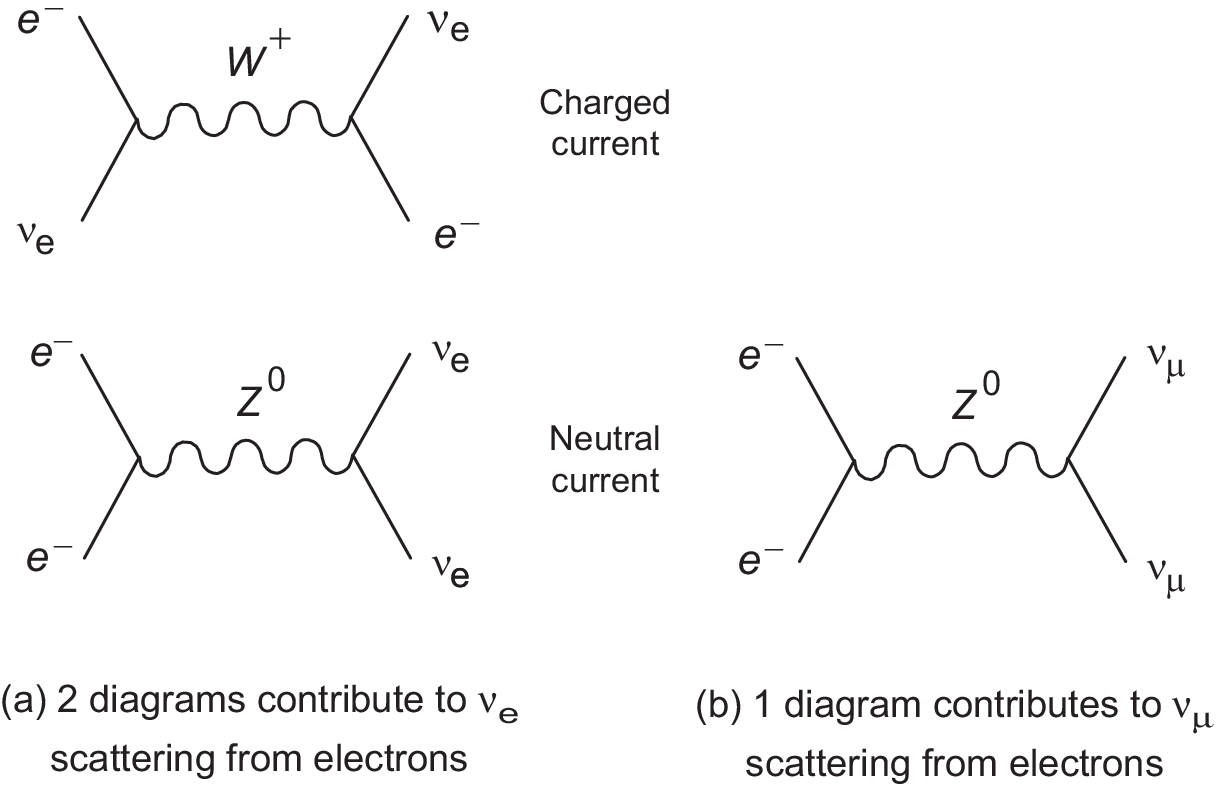}
{0pt}
{0pt}
{0.68}
{Feynman diagrams  responsible for neutrino--electron scattering in a 2-flavor 
model with $\nu\tsub e$ and $\nu_\mu$ flavors. Such diagrams are pictorial 
representations of quantum-mechanical matrix elements.  For example, the upper 
left diagram stands for the matrix element ($\sim$ probability amplitude)
for a process where  an electron neutrino exchanges a virtual $W^+$ 
intermediate vector boson with an electron, with the neutrino converted to an 
electron and the electron converted to an electron neutrino.  This is a {\em 
charged-current} process, because the $W^+$ virtual exchange particle is 
electrically charged. On the other hand, the lower left diagram describes a {\em neutral-current} process where the exchanged virtual $Z^0$ carries no electrical charge.  Electron neutrinos can interact with electrons through both the charged-current and neutral-current diagrams shown in (a), but muon and tau neutrinos can interact with electrons only through the neutral-current diagram in (b).}

The neutral current interaction contributes to both electron and muon neutrino 
scattering, so let's ignore it and concentrate on the charged-current diagram in 
\fig{MSWFeynman}(a) that contributes only to $\nu\tsub e$ elastic scattering in 
a 2-flavor model. From standard methods of quantum field theory 
\cite{kuo1989,smir2003,blen2013} the charged-current diagram in \fig{MSWFeynman} 
contributes a medium-dependent interaction having the form of an 
effective potential energy $V$ seen {\em only by the electron neutrinos or 
antineutrinos}, with a magnitude
\begin{equation}
   V = \pm \sqrt 2 G\tsub F n\tsub e ,
   \label{nosc14}
\end{equation} 
where the positive sign is for electron neutrinos (our primary concern here), 
the negative sign is for electron antineutrinos, $n\tsub e$ is the local 
electron number density, and $G\tsub F$ is the weak (Fermi) coupling constant characterizing the strength of the weak interactions.
%
%\subsubsection{\label{effectivenmass} The Effective Neutrino Mass in Medium}
%
The energy is given by $E^2 = p^2 + m^2$.
For the electron 
neutrino subject to the additional potential $V$
\begin{equation}
p^2 + m^2 = (E-V)^2 = E^2 \left(1-\frac VE\right)^2
\simeq  E^2 - 2EV,
\end{equation}
where the last step is justified by assuming that $V\muchless E$. Thus 
\begin{equation}
 E^2 \sim p^2 + (m^2 + 2EV).   
\end{equation}
Since $V$ is positive, an electron neutrino 
behaves effectively as if it is slightly heavier when propagating through 
matter 
than in vacuum, with the amount of increase governed by the electron density of 
the matter, but a muon or tau neutrino is unaffected because they do not see the effective potential $V$. \fig{effectiveNeutrinoMass} illustrates.%
\singlefig
{effectiveNeutrinoMass}
{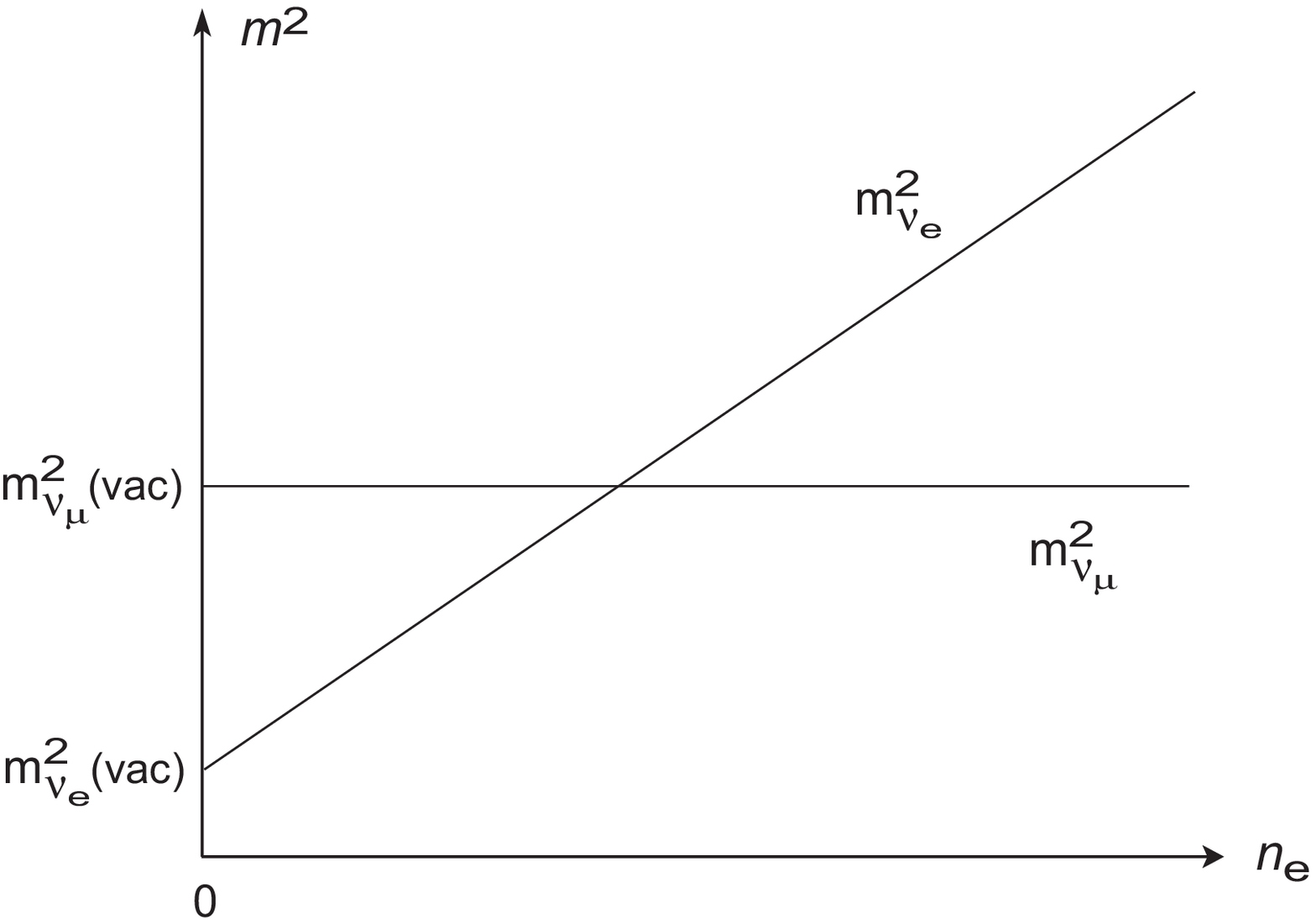}
{0pt}
{0pt}
{0.49}
{The effective mass-squared  of electron neutrinos and muon neutrinos as a 
function of electron number density $n\tsub e$, neglecting flavor mixing. 
Because the $\nu_\mu$ does not couple to the charged weak current its $m^2$ does 
not depend on $n\tsub e$ but the effective $m^2$ of $\nu\tsub e$ increases 
linearly with the electron density. Thus the $m^2$ spectrum in vacuum can be 
{\em inverted in matter} at high electron density.
}
From this figure, an electron neutrino that is less massive than a muon neutrino 
in vacuum will become effectively {\em more 
massive} than its oscillation partner in matter if the electron density is 
sufficiently high.

\subsection{\label{sna1.7.1} The Mass Matrix}

To address in more depth neutrino oscillations in matter, let's introduce 
a more formal derivation of the neutrino oscillation problem that will prove to be useful for 
subsequent discussion. \cite{bah89,wol78,bet86,kuo1989,col89b,mori2004}

\subsubsection{\label{LHprop} Propagation of Left-Handed Neutrinos}

First note that the full spin structure does not influence the propagation of 
neutrinos in the absence of magnetic fields because  only the 
left-handed component of the neutrino couples to the weak interactions. The 
Schr\"odinger equation of ordinary quantum mechanics is not relativistically 
invariant and so is not appropriate for relativistic particles. For relativistic
fermions the wave equation must be generalized to the {\em Dirac equation}, 
while for spinless particles the corresponding relativistic wave equation is the 
{\em Klein--Gordon equation.}  
Relativistic fermions are  generally described by a Dirac equation 
but if the spin structure is eliminated from consideration the propagation 
of a (left-handed component of the) free neutrino may instead be described by 
the simpler free-particle Klein--Gordon equation
\begin{equation}
(\dalem +m^2)\ket\nu = 0
\qquad
\dalem \equiv -\frac{\partial^2}{\partial t^2} 
+\frac{\partial^2}{\partial x^2}
+\frac{\partial^2}{\partial y^2}
+\frac{\partial^2}{\partial z^2} ,
\label{nmm1.1}
\end{equation}
 where $\dalem$ is termed the {\em d'Alembertian operator} and for $n$ neutrino 
flavors $\ket\nu$ is an $n$-component column vector in the mass-eigenstate basis 
and $m^2$ is an $n\times n$ matrix.

Because of oscillations, the solutions 
to \eq{nmm1.1} of interest correspond to the propagation of a linear 
combination of mass 
eigenstates. For ultrarelativistic neutrinos one makes only small errors by 
assuming neutrinos of tiny mass and slightly different energies to propagate 
with the same 3-momentum $\bm p$. In that approximation
a solution of \eq{nmm1.1} for definite momentum is given by
\begin{equation}
\ket{\nu_i} = e^{-iE_i t} \cdot e^{-i\sprod{p}{x}}
\qquad
E_i = \sqrt{\bm p^2 + m_i^2} .
\label{nmm1.2}
\end{equation}
For ultrarelativistic particles $|\bm p| \muchgreater m_i$ so
\begin{equation}
E_i = \left[\bm p^2 \left(1+\frac{m_i^2}{\bm p^2}\right)\right]^{1/2}
\simeq \,\bm p + \frac{m_i^2}{2|\bm p|}
\simeq \,\bm p + \frac{m_i^2}{2 E} ,
\label{nmm1.2b}
\end{equation}
where $E \sim p \equiv |\bm p|$.
Thus, from \eq{nmm1.2}
$$
\ket{\nu_i(t)} 
\,\simeq\, e^{-i|\bm p|t} \cdot e^{-i\sprod{p}{x}} e^{-i\bm p \cdot \bm x}
\cdot e^{-i (m_i^2/2E) t}
.
$$
But the initial exponential factors involving the 3-momenta provide only a 
common overall phase 
that does not affect observables, so they may be dropped to give
\begin{equation}
\ket{\nu_i(t)}
\simeq\, e^{-i (m_i^2/2E) t} .
\label{nmm1.3}
\end{equation}
 Differentiating \eq{nmm1.3} with respect to 
time gives an equation of motion for a single mass eigenstate labeled by $i$,
\begin{equation}
i\, \frac{d}{dt}\,\ket{\nu_i(t)} = \frac{m_i^2}{2E} \,\ket{\nu_i(t)},
\label{nmm1.4}
\end{equation}
which may be generalized for a 2-flavor model to the matrix equation%
\begin{equation}
i \,\frac{d}{dt}  \onematrix{\nu_1}{\nu_2} =
\twomatrix{m_1^2/2E}{0}{0}{m_2^2/2E} 
\onematrix{\nu_1}{\nu_2} ,
\label{nmm1.5}
\end{equation}
where the quantity
$$
M  \equiv
\twomatrix{m_1^2/2E}{0}{0}{m_2^2/2E}
$$
is termed the {\em mass matrix.} 
In these equations elapsed time $t$ or distance traveled $r$ may 
be used interchangeably as the independent variable, since $r\sim ct$ for 
ultrarelativistic neutrinos.

\subsubsection{Evolution in the Flavor Basis}

Neutrinos propagate in mass eigenstates but they are produced and detected in 
flavor eigenstates, so it is useful to express the preceding equation in 
the flavor basis. The transformations matrices are given in Eqs.\ 
(\ref{2flavorVac}) and (\ref{2massVac}), which permits \eq{nmm1.5} to be 
written in the 
form
\begin{align*}
i \,\frac{d}{dt}  \onematrix{\nu_1}{\nu_2} &=
i \,\frac{d}{dt} \, U^\dagger\onematrix{\nu\tsub e}{\nu_\mu}
\\
&=
\twomatrix{m_1^2/2E}{0}{0}{m_2^2/2E} U^{\dagger}
\onematrix{\nu\tsub e}{\nu_\mu} .
\end{align*}
Multiplying from the left by $U$ and using the unitarity condition $UU^\dagger 
= 1$ gives
\begin{equation}
i \,\frac{d}{dt}  \onematrix{\nu\tsub e}{\nu_\mu} =
U \twomatrix{m_1^2/2E}{0}{0}{m_2^2/2E} U^{\dagger}
\onematrix{\nu\tsub e}{\nu_\mu} ,
\label{nmm1.6}
\end{equation}
where the flavor and mass eigenstates are related by
$$
\onematrix{\nu\tsub e}{\nu_\mu} = U 
\onematrix{\nu_1}{\nu_2} 
\qquad
\onematrix{\nu_1}{\nu_2} = U^\dagger \onematrix{\nu\tsub e}{\nu_\mu}
,
$$
and the transformation matrices between mass and flavor bases may be written explicitly as
$$
 U = \twomatrix{\cos\thetav}{\sin\thetav}{-\sin\thetav}{\cos\thetav}
\qquad U^\dagger = 
\twomatrix{\cos\thetav}{-\sin\thetav}{\sin\thetav}{\cos\thetav}.
$$
As is clear by substitution, \eq{nmm1.6}  has a solution
$$
\onematrix{\nu\tsub e(t)}{\nu_\mu(t)} =
U \twomatrix{e^{-i(m_1^2/2E) t}}
{0}{0}
{e^{-i(m_2^2/2E) t}} U^{\dagger}
\onematrix{\nu\tsub e(0)}{\nu_\mu(0)} .
%\label{nmm1.7}
$$
Since no individual neutrino mass has been measured thus far,
it is convenient to rewrite \eq{nmm1.6} in terms of $\Delta m^2$, which has been measured.
Adding a multiple of the unit matrix to the matrix in \eq{nmm1.6} 
will not modify quantum observables (a trick that will be employed several 
times in what follows), so let's subtract $m_1^2/2E$  times the unit $2\times 
2$ matrix from the matrix in \eq{nmm1.6} 
and use
$$
\twomatrix{0}{0}{0}{\Delta m^2/2E} =
 \twomatrix{m_1^2/2E}{0}{0}{m_2^2/2E} 
-
\twomatrix{m_1^2/2E}{0}{0}{m_1^2/2E}
$$
to replace \eq{nmm1.6} by the equivalent form
\begin{equation}
i \,\frac{d}{dt}  \onematrix{\nu\tsub e}{\nu_\mu} =
U \twomatrix{0}{0}{0}{\Delta m^2/2E} U^{\dagger}
\onematrix{\nu\tsub e}{\nu_\mu} ,
\label{nmm1.6b}
\end{equation}
where the mass-squared difference $\Delta m^2 \equiv m_2^2 - m_1^2$ was introduced in \eq{vosc6}.

\subsubsection{Propagation in Matter}

The flavor evolution equation (\ref{nmm1.6b}) is just a reformulation of our previous 
treatment of neutrinos propagating in vacuum, so it contains nothing new.  However,  let us now add a charged-current 
interaction with matter.  By previous arguments the charged 
current couples only elastically and only to electron neutrinos, so we add 
to 
the Klein--Gordon equation (\ref{nmm1.1}) an interaction potential given by 
\eq{nosc14}, which modifies the equation of motion (\ref{nmm1.6b}) to
$$
i \,\frac{d}{dt}  \onematrix{\nu\tsub e}{\nu_\mu} =
\left[ U \twomatrix{0}{0}{0}{\Delta m^2/2E} U^{\dagger}
+\twomatrix{V(t)}{0}{0}{0} \right]
\onematrix{\nu\tsub e}{\nu_\mu} ,
%\label{matter1.1}
$$
where  the effective potential generated by the charged-current coupling of the electron neutrino to the matter,
\begin{equation}
V(t) = \sqrt 2 G\tsub F \, n\tsub e(t) ,
\end{equation}
depends on time (or equivalently on position),
because it is a function of the electron density. Inserting the explicit forms for the 
transformation matrices $U$ and $U^\dagger$ gives
\begin{align}
&i \,\frac{d}{dt}  \onematrix{\nu\tsub e}{\nu_\mu} =
\nonumber
\\
&\left[
\U
\twomatrix{0}{0}{0}{\Delta m^2/2E}
\Udag
\right.
\nonumber
\\
 &\left. \quad+ 
\twomatrix{V}{0}{0}{0} \right] \onematrix{\nu\tsub e}{\nu_\mu}
\nonumber
\\
&=
\left[
\begin{pmatrix}
{\dfrac{\Delta m^2}{2E} \sin^2\thetav} &
{\dfrac{\Delta m^2}{2E} \sin\thetav \cos\thetav}
\\[6pt]
{\dfrac{\Delta m^2}{2E} \sin\thetav \cos\thetav} &
{\dfrac{\Delta m^2}{2E} \cos^2 \thetav}
\end{pmatrix}
+
\twomatrix{V}{0}{0}{0}
\right]
\onematrix{\nu\tsub e}{\nu_\mu}
\nonumber
\\
&=\left[
\frac{\Delta m^2}{2E}
\twomatrix
{0}
{\sin\thetav \cos\thetav}
{\sin\thetav \cos\thetav}
{\cos^2\thetav - \sin^2\thetav}
+
\twomatrix{V}{0}{0}{0}
\right]
\onematrix{\nu\tsub e}{\nu_\mu}
\nonumber
\\
&=
\begin{pmatrix}
{V} &
{\dfrac{\Delta m^2}{4E} \sin2\thetav} 
\\[6pt]
{\dfrac{\Delta m^2}{4E} \sin2\thetav} &
{\dfrac{\Delta m^2}{2E} \cos2\thetav}
\end{pmatrix}
\onematrix{\nu\tsub e}{\nu_\mu} 
\equiv
M \onematrix{\nu\tsub e}{\nu_\mu},
\label{matter1.2}
\end{align}
where in the second step 
$
(\Delta m^2/4E)\sin^2\thetav 
$
times the $2\times 2$ unit matrix has been subtracted, in the final line the 
trigonometric identities
$$
\sin\thetav \cos\thetav = \tfrac12 \sin 2\thetav
\qquad
\cos^2\thetav - \sin^2\thetav = \cos2\thetav
$$
have been used, and $M$ is termed the {\em mass matrix} (in the flavor basis).

\eq{matter1.2} is the required result but it is conventional to 
write the mass matrix $M$ appearing in it in a more symmetric form. 
First define
\begin{equation}
A \equiv 2EV = 2\sqrt 2 EG\tsub F n\tsub e,
\label{matter1.3}
\end{equation}
(which has units of mass squared) and then subtract 
$
A/4E + (\Delta m^2/4E) \cos2\thetav
$
% $
% [\tfrac14 AE + (\Delta m^2/4E) \cos2\thetav]
% $
multiplied by the unit matrix
to give the mass matrix in traceless form
\begin{equation}
\begin{split}
M &= 
\frac{\Delta m^2}{4E}
\twomatrix
{\dfrac{A}{\Delta m^2} - \cos2\thetav}
{\sin2\thetav}
{\sin2\thetav}
{\cos2\thetav - \dfrac{A}{\Delta m^2}}
\\
&= \displaystyle\frac{\pi}{L} \twomatrix
         {\chi - \cos 2\theta}
         {\sin 2\theta}
         {\sin 2\theta}
         {\cos 2\theta -\chi},  
\label{nosc15}
\end{split}
\end{equation}
where the dimensionless charged-current coupling strength  $\chi$ has been introduced 
through
\begin{equation}
\chi \equiv \frac{L}{\ell\tsub m} = \frac{2EV}{\Delta m^2} 
\qquad
\ell \tsub m \equiv \frac{\sqrt 2 \pi}{G\tsub F n\tsub e}
\qquad
L \equiv \frac{4\pi E}{\Delta m^2} ,
\label{nosc16b}
\end{equation}
with $L$ the vacuum oscillation length defined in \eq{vosc8} and $\ell \tsub m$ 
an additional contribution to the oscillation length in matter that is termed 
the {\em refraction length} because it is a characteristic distance over which 
there is significant refraction of the neutrino by the matter. \cite{smir2003}

\subsection {\label{matterSolutions1.1} Solutions in Matter}

For a fixed density the mass eigenstates in matter---which generally will differ 
from the mass eigenstates in vacuum because of the interaction $V$---may be 
found by diagonalizing (finding the eigenvalues) of the mass matrix $M$ at that 
density. However, since the interaction $V$ depends on the density, in a medium 
with varying density such as the Sun the mass eigenstates in matter at one time 
(or position) will generally not be eigenstates at another time. Let's imagine 
dividing the Sun up into concentric radial layers,  as illustrated in 
\fig{solarLayers},%
\singlefig
{solarLayers} 
{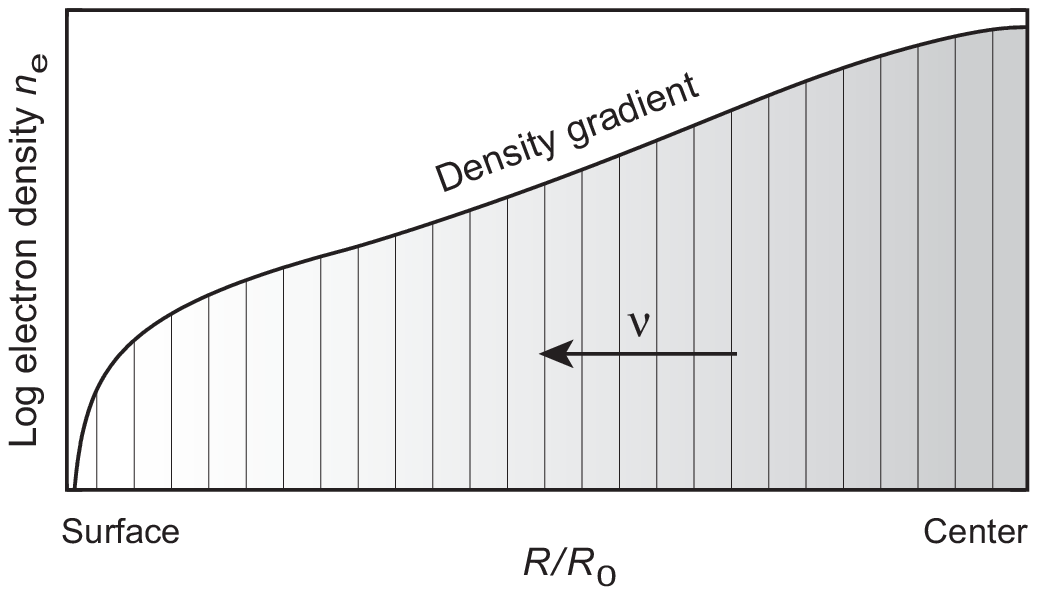}
{0pt}
{0pt}
{0.72}
{Solar density gradient. Neutrinos are produced near the 
center at high density and propagate out through regions of
decreasing density. Within a given concentric layer, the 
density may be assumed to be nearly constant.} 
and assume the density within each layer to be approximately constant.
Our approach will be to first understand how to calculate the mass 
eigenstates within a single layer assumed to have a constant density, and then 
to determine the evolution of neutrino states as they 
propagate through successive layers of decreasing density in the Sun.

\subsubsection {Mass Eigenvalues for Constant Density}

{\em At constant density} the problem resembles vacuum oscillations, except with 
a different potential $V \ne 0$, and the time-evolved mass states in matter, 
$\mattermass1$ and $\mattermass2$, may be found by determining the eigenvalues 
of the mass matrix (\ref{nosc15}).  The eigenvalues of an $n\times n$ matrix 
$A$ correspond to the values of $\lambda$ that solve the characteristic equation
$
\det(A-\lambda I) = 0,
$
with $I$ the $n\times n$ unit matrix and $\det$ indicating the determinant.  For 
a  general $2\times 2$ matrix
$$
A = 
\begin{pmatrix}
\alpha & \beta
\\
\gamma & \delta
\end{pmatrix} ,
$$
the characteristic equation is quadratic with the two solutions,
$$
\lambda_\pm = \tfrac12 \trace(A) \pm \sqrt{\tfrac14 \trace(A)^2 - \det(A)},
$$
where $\trace A$ is the trace and $\det A$ is the 
determinant of $A$.
Hence the eigenvalues $\lambda_\pm$  of the mass matrix (\ref{nosc15}) are given by
\begin{equation}
\begin{split}
\lambda_\pm &= \left(\frac{m_1^2 + m_2^2}{2} + \frac{\Delta m^2}{2} \chi 
\right)
\\
 &\quad\pm \frac{\Delta m^2}{2}
\sqrt{ (\cos2\thetav - \chi)^2 + \sin^2 2\thetav}.
\end{split}
\label{matter1.4}
\end{equation}
The second term gives the splitting between  the two mass eigenstates and is minimal at 
the density where $\chi = \cos2\thetav$. By analogy with the vacuum case, the 
mass eigenstates in matter, $\nu_1^{\rm m}\equiv \mattermass1$ and $\nu_2^{\rm m}\equiv\mattermass2$ at  fixed time 
$t$, are assumed to be related to the flavor eigenstates by
\begin{equation}
 \onematrix
{\nu\tsub e}
{\nu_\mu}
=
U\tsub m(t) 
\onematrix
{\nu_1^{\rm m}}
{\nu_2^{\rm m}} .
\label{matter1.5}
\end{equation}
where $U\tsub m(t)$ is a unitary matrix depending on the density that we must now determine.

\subsubsection{The Matter Mixing Angle $\bm\thetam$}

The matrix $U\tsub m(t)$ can be parameterized as for the 
vacuum mixing matrix $U$, but now in terms of a time-dependent matter mixing angle 
$\thetam(t)$ with
\begin{equation}
 U\tsub m =
\twomatrix
{\cos\thetam}
{\sin\thetam}
{-\sin\thetam}
{\cos\thetam}
\quad
 U_{\rm\scriptstyle m}^\dagger =
\twomatrix
{\cos\thetam}
{-\sin\thetam}
{\sin\thetam}
{\cos\thetam} .
\label{matter1.5b}
\end{equation}
The relationship of the matter mixing angle $\thetam$ and vacuum mixing angle 
$\thetav$ at time $t$ can be established by requiring that a similarity 
transform  by the unitary matrix $U\tsub m(t)$ diagonalize the mass matrix, 
with the diagonal 
elements being the time-dependent eigenvalues in matter $E_1(t)$ and $E_2(t)$,
\begin{equation}
U_{\rm\scriptstyle m}^\dagger (t) M U\tsub m(t) = \diag [E_1(t), E_2(t)]
=
\twomatrix
{E_1(t)}
{0}
{0}
{E_2(t)} .
\label{matter1.6}
\end{equation}
Inserting the explicit values of the 
matrices $U\tsub m$, $U^\dagger_{\rm m}$, and $M$ from Eqs.\ (\ref{nosc15}) and 
(\ref{matter1.5b}) in \eq{matter1.6} gives a matrix equation, for which comparison of matrix entries  on the two sides of the equation requires the  
matter and vacuum mixing angles to be related by
\begin{equation}
   \tan 2\thetam = \frac{\sin2\thetav}{\cos2\thetav
       \pm \chi}
 = \frac{\tan 2\thetav}{1\pm \chi/\cos 2\thetav} ,
   \label{nosc18}
\end{equation}
where the plus sign is for $m_1 > m_2$ and the negative sign for $m_1 < m_2$. 
From \eq{nosc18},  in vacuum $\thetam = \thetav$  but in matter the mixing angle 
will be modified from its vacuum value by a density-dependent amount governed by the charged-current coupling strength $\chi$.

\subsubsection{The Matter Oscillation Length $\bm L\tsub {m}$}

From \eq{vosc8} the vacuum oscillation length is proportional to $1/\Delta m^2$. 
In matter the $\nu\tsub e$ effective mass is altered by interaction with the 
medium and the vacuum mass-squared difference is rescaled, 
$
\Delta m^2 \rightarrow f(\chi) \Delta m^2 ,
$
where from the splitting of the two eigenvalues in \eq{matter1.4}
\begin{align}
f(\chi) &= \sqrt{(\cos2\thetav - \chi)^2 + \sin^2 2\thetav}
\nonumber
\\
&= \sqrt{1 - 2\chi \cos2\thetav + \chi^2} .
\label{matter1.7}
\end{align}
Hence the oscillation length in matter $L\tsub m$ is 
given by
\begin{align}
   L\tsub m &= \frac{4\pi E}{f(\chi) \Delta m^2} 
= \frac{L}{f(\chi)}
\nonumber
\\
&=
\frac{L}{\sqrt{(\cos 
2\thetav - \chi)^2
       + \sin^2 2\thetav}} ,
   \label{nosc19}
\end{align}
which reduces to the vacuum oscillation length $L$ if $\chi \rightarrow 0$. The 
variations of $\thetam$, $L\tsub m$, and $f$ with the dimensionless 
charged-current coupling strength $\chi$ are illustrated in 
\fig{matterAngleLength}%
\singlefig
{matterAngleLength}
{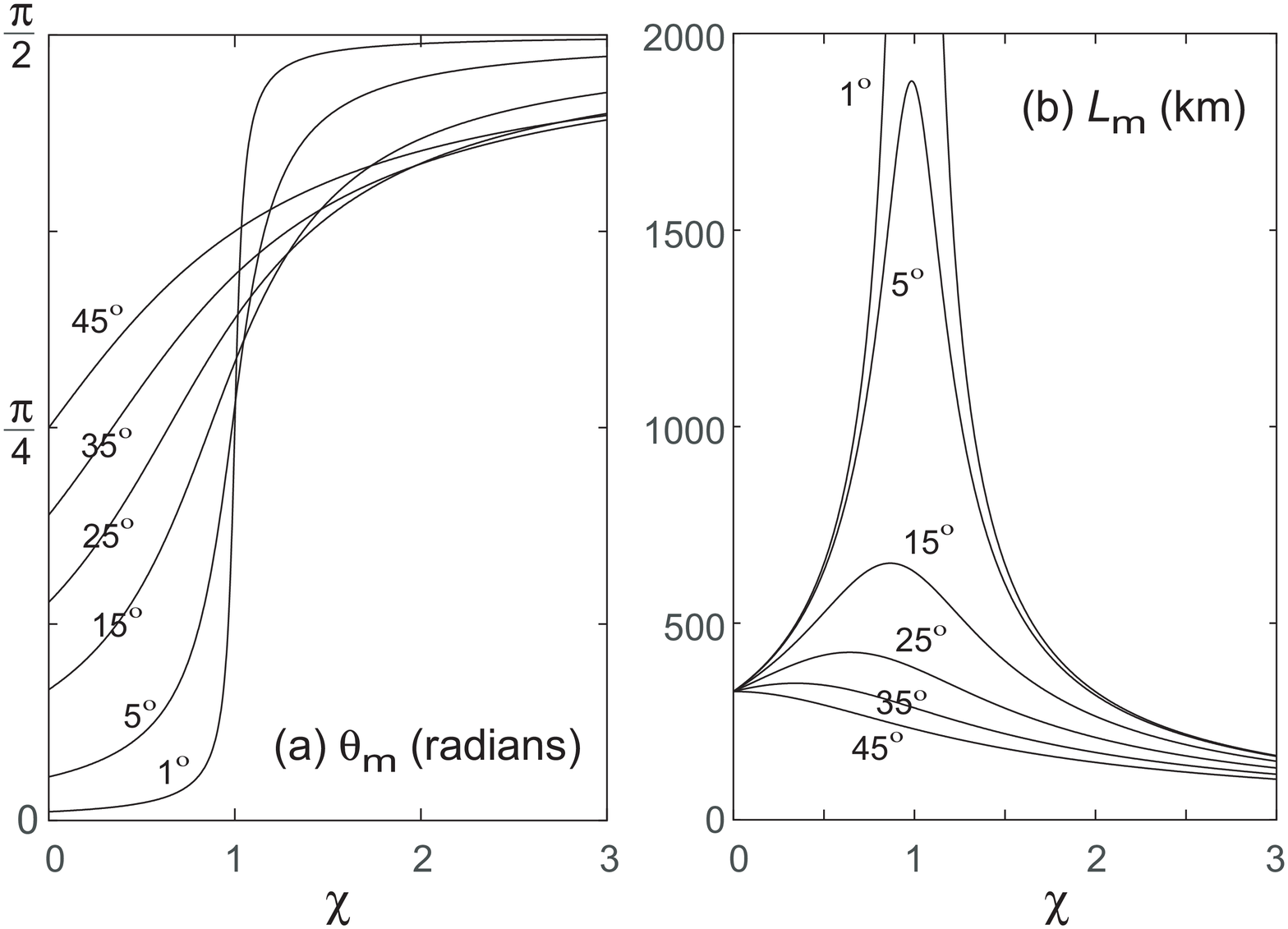}
{0pt}
{0pt}
{0.297}
{(a)~Mixing angle in matter $\thetam(\chi)$ and (b)~oscillation length in 
matter $L\tsub{m}(\chi)$,  as a function of the dimensionless matter coupling parameter 
$\chi$.  All calculations assumed $E=10\units{MeV}$ and $\Delta m^2 = 7.6\times 
10^{-5} \punits{eV}{2}$, and curves are marked with the assumed
vacuum mixing angle $\thetav$.
}
for several values of the vacuum mixing angle $\thetav$. 

From \fig{matterAngleLength}(a), the matter mixing angle $\thetam$ reduces to 
the vacuum mixing angle $\thetav$ as $\chi \rightarrow 0$, but $\thetam 
\rightarrow \frac\pi2$ for {\em any value of the vacuum mixing angle} as $\chi 
\rightarrow \infty$, with the rate of approach to $\tfrac\pi2$ being fastest for 
smaller vacuum angle. From \fig{matterAngleLength}(b) the matter oscillation 
length is equal to the vacuum oscillation length at zero coupling, but increases 
to a maximum at the coupling strength where $\thetam = \frac\pi4$ [compare 
\fig{matterAngleLength}(a)], and then decreases again.  The coupling strength at 
which $L\tsub m$ is maximal coincides with highest rate of change in $\thetam$, 
and with the minimum of the scaling function $f(\chi)$ displayed in 
\fig{matterAngleLength}(c). The rapid change of $\thetam$ and the strong peaking 
of $L\tsub m$ near the density where $\thetam = \frac\pi4$  are suggestive of 
{\em resonant behavior,}  as will be elaborated shortly.

\subsubsection{Flavor Conversion Probabilities in Constant-Density Matter}

In  {\em constant-density matter} the electron neutrino state after a 
time $t$ becomes 
\begin{equation}
\begin{split}
      \ket{\nu(t)} = (\cos^2\thetam e^{-iE_1t}
      + \sin^2\thetam e^{-iE_2t}) \ket{\nu\tsub e}
\\
      + \sin\thetam
          \cos\thetam (-e^{-iE_1t} + e^{-iE_2t}) \ket{\nu_\mu},
\end{split}
   \label{nosc17}
\end{equation}
which is analogous to the vacuum equation \eq{vosc3} with  $\thetav$ replaced by 
 $\thetam$ defined through 
\eq{nosc18}. The corresponding flavor 
conservation and retention probabilities are given by \eq{flavorProbs} with the 
replacements $\thetav \rightarrow \thetam$ and $L \rightarrow L\tsub m$,
\begin{equation}
\begin{array}{c}
\pee r 
      = 1 - \sin^2 2\thetam \sin^2 \left( \displaystyle
\frac{\pi r}{L\tsub m} \right)
\end{array}
\label{flavorProbsMSW}
\end{equation}
with $\pemu r = 1- \pee r$.  The corresponding 
classical averages are
\begin{equation}
\peec =
 1 - \tfrac12 \sin^2 2\thetam
\quad
\pemuc =
 \tfrac12 \sin^2 2\thetam ,
\label{flavorProbsMSWavg}
\end{equation}
which are appropriate when the uncertainty in distance between source and 
detection exceeds the oscillation length (which is generally the case for solar 
neutrinos).

\subsection{\label{sna1.7.3} The MSW Resonance Condition}

From \eq{flavorProbsMSW}, optimal flavor mixing occurs whenever 
$\sin^2 2\thetam \rightarrow 1$, implying that
$|\thetam| = \frac{\pi}{4}$.  The most significant property of \eq{nosc18} 
relative to the vacuum solution is that if $\Delta m^2$ and $L$ are positive 
[requiring that $m_1 < m_2$, which selects the negative sign in 
\eq{nosc18}]. 
Then $\tan2\thetam \rightarrow 
\pm\infinity$ and $\thetam \rightarrow \frac{\pi}{4}$ whenever 
\begin{equation}
    \chi = \cos 2\thetav,
   \label{nosc20}
\end{equation} 
where $\chi = L/\ell\tsub m$,
as illustrated in \fig{resonanceMSW}.%
\singlefig
{resonanceMSW} 
{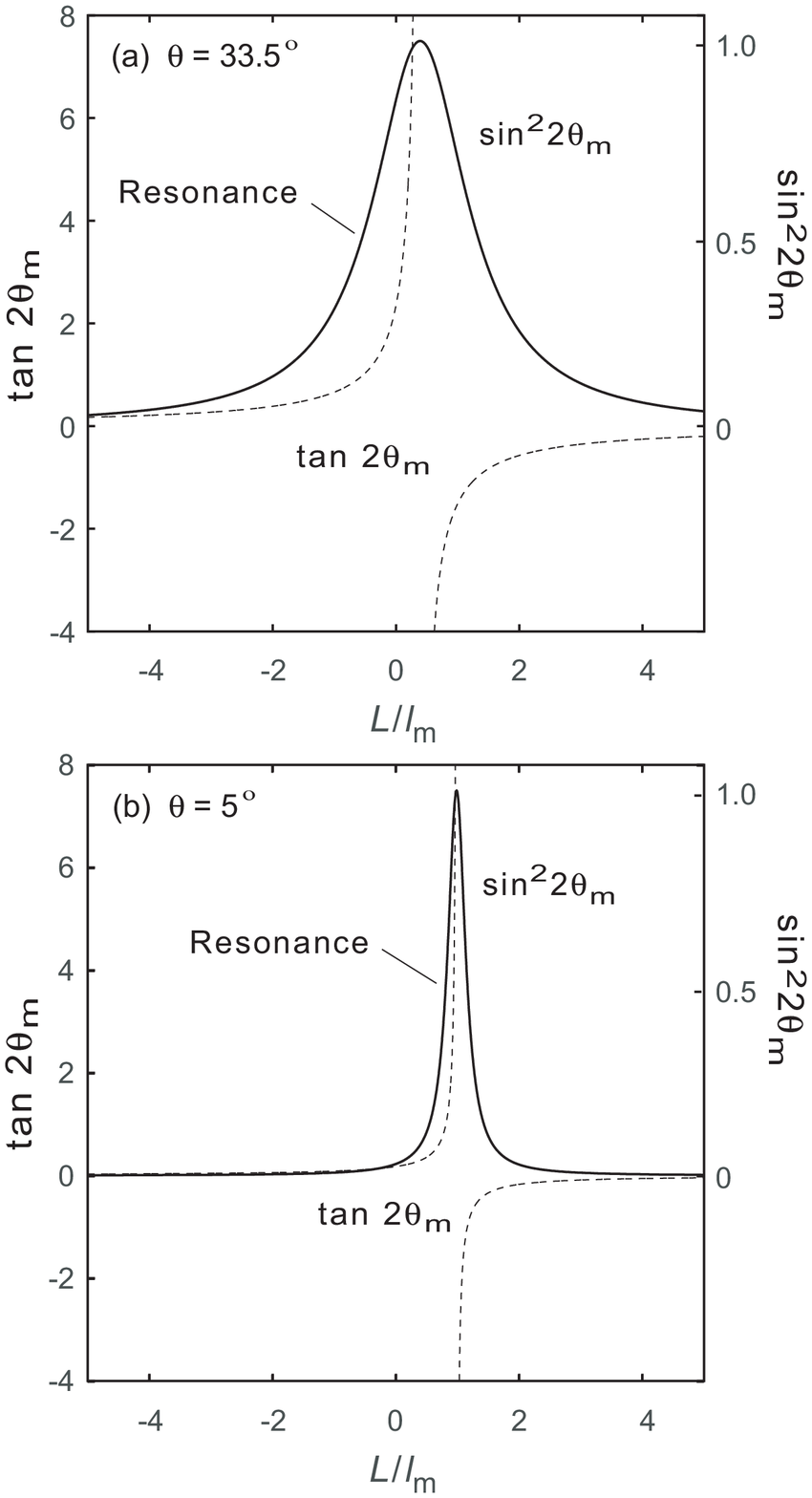}
%{figures/resonanceMSW2.eps}
{0pt}
{0pt}
{0.54}
{The MSW resonance condition for two values of the vacuum mixing angle 
$\thetav$. When $\chi = L/\ell\tsub m \rightarrow \cos 2\thetav$ the denominator of 
\eq{nosc18} goes to zero, $\tan 2\thetam$ goes to $\pm \infty$ so that 
$|\thetam| \rightarrow \frac\pi4$, and the flavor conversion probability $\sin^2 
2\thetam$ attains its maximum value. Thus, at the resonance \eq{flavorProbsMSW} 
indicates that it is possible to obtain very large flavor conversion  for any 
non-vanishing vacuum oscillation angle $\thetav$.}
From  \eq{nosc16b},  this occurs when the 
electron density satisfies
\begin{equation}
   n\tsub e = \frac{\cos2\thetav \Delta m^2}
              {2\sqrt 2 G\tsub F E} \equiv \resonanceDensity.
   \label{nosc21}
\end{equation}
From \eq{flavorProbsMSW}, this corresponds to a resonance condition leading to 
maximal flavor mixing between electron neutrinos and muon neutrinos, with a 
$\nu\tsub 
e$ survival probability
\begin{equation}
\pee r =
 1-\sin^2
         \left( \frac{\pi r}{L\tsub m} \right) \qquad {\rm (at\ resonance)} ,
   \label{nosc22}
\end{equation}
and, from Eqs.\ (\ref{nosc19}) and  (\ref{nosc20}), an oscillation length at 
resonance $L_{\scriptscriptstyle\rm m}^{\rm\scriptscriptstyle R}$ given by
\begin{equation}
L_{\scriptscriptstyle\rm m}^{\rm\scriptscriptstyle R}=
   L\tsub m (\chi \hspace{-2pt}=\hspace{-2pt} \cos 2\theta) 
= \frac{L}{\sin2\thetav} .
%\qquad {\rm (Resonance)} .
   \label{nosc23}
\end{equation}
This is the {\em Mikheyev--Smirnov--Wolfenstein} or {\em MSW resonance}. 
\cite{wol78,mik86} It implies that---no matter how small the vacuum mixing 
angle $\thetav$---as long as it is not zero there is some critical value 
\resonanceDensity\ of the electron density defined by \eq{nosc21} where the 
resonance condition is satisfied and {\em maximal flavor mixing} ensues. The 
important resonance parameters are plotted in \fig{MSWparameters_ne}%
\singlefig
{MSWparameters_ne} 
{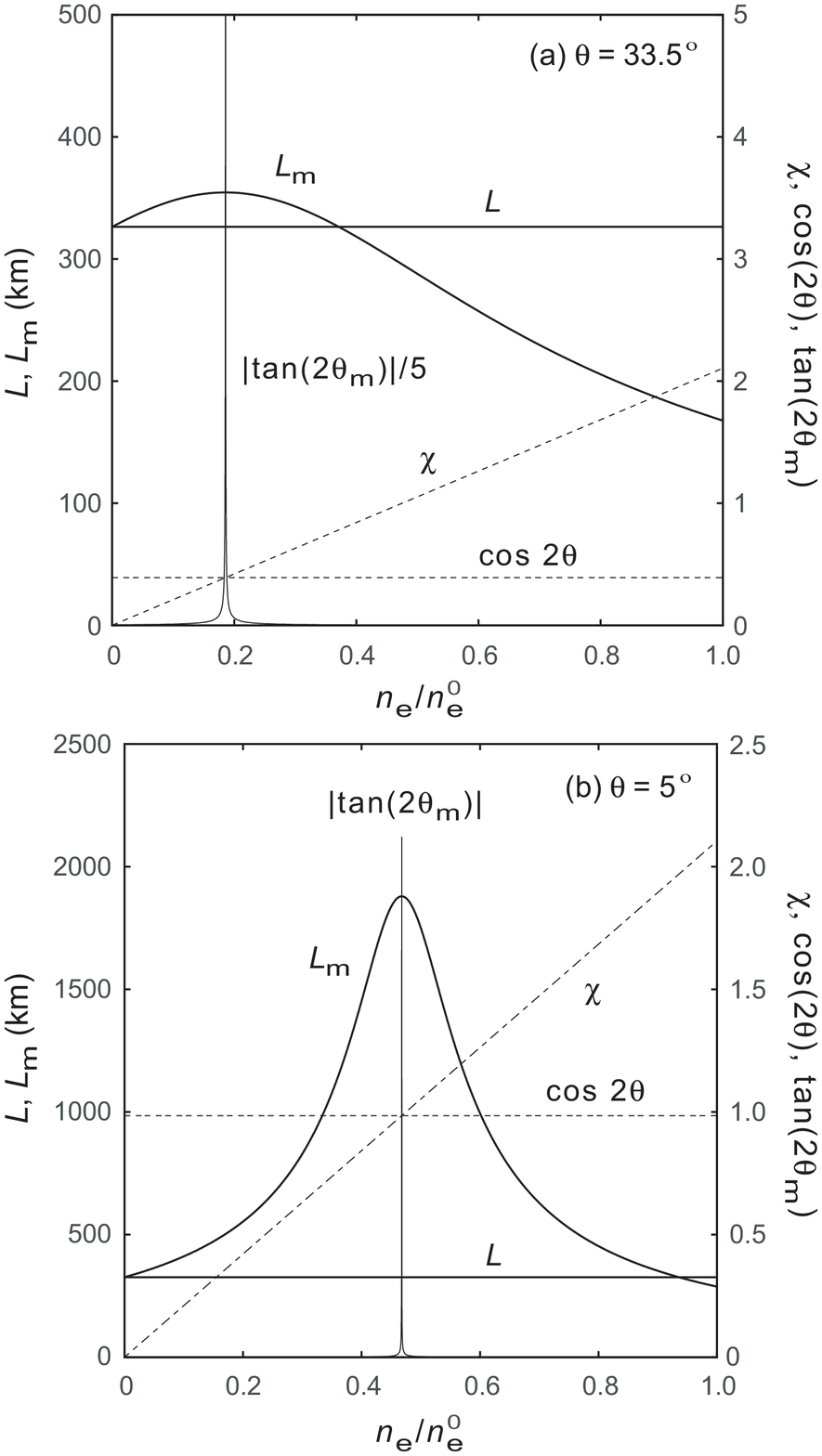}
%{figures/MSWparameters_ne.eps}
{0pt}
{0pt}
{0.47}
{Resonance parameters versus the electron number density $n\tsub e$ in 
units of the central solar value $n_{\rm\scriptstyle e}^0 \sim 6.3 \times 
10^{25} \punits{cm}{-3}$ for  $\theta = 33.5^\circ$ and $5^\circ$, with $\Delta 
m^2 = 7.6\times10^{-5} \punits{eV}{2}$ and $E=10\punits{MeV}{}$. The coupling 
strength  $\chi = L/\ell\tsub m$  is linear in the density. Intersection of 
the dashed lines specifies the density giving the resonance condition.  
}
as a function of electron density for two values of the vacuum mixing angle 
$\theta$. 

The effect of the MSW resonance on variation of the matter mixing angle 
$\thetam$ and the oscillation length in matter $L\tsub m$ are illustrated for a 
small and large angle solution in \fig{matterAngle_composite}.%
\singlefig
{matterAngle_composite} 
{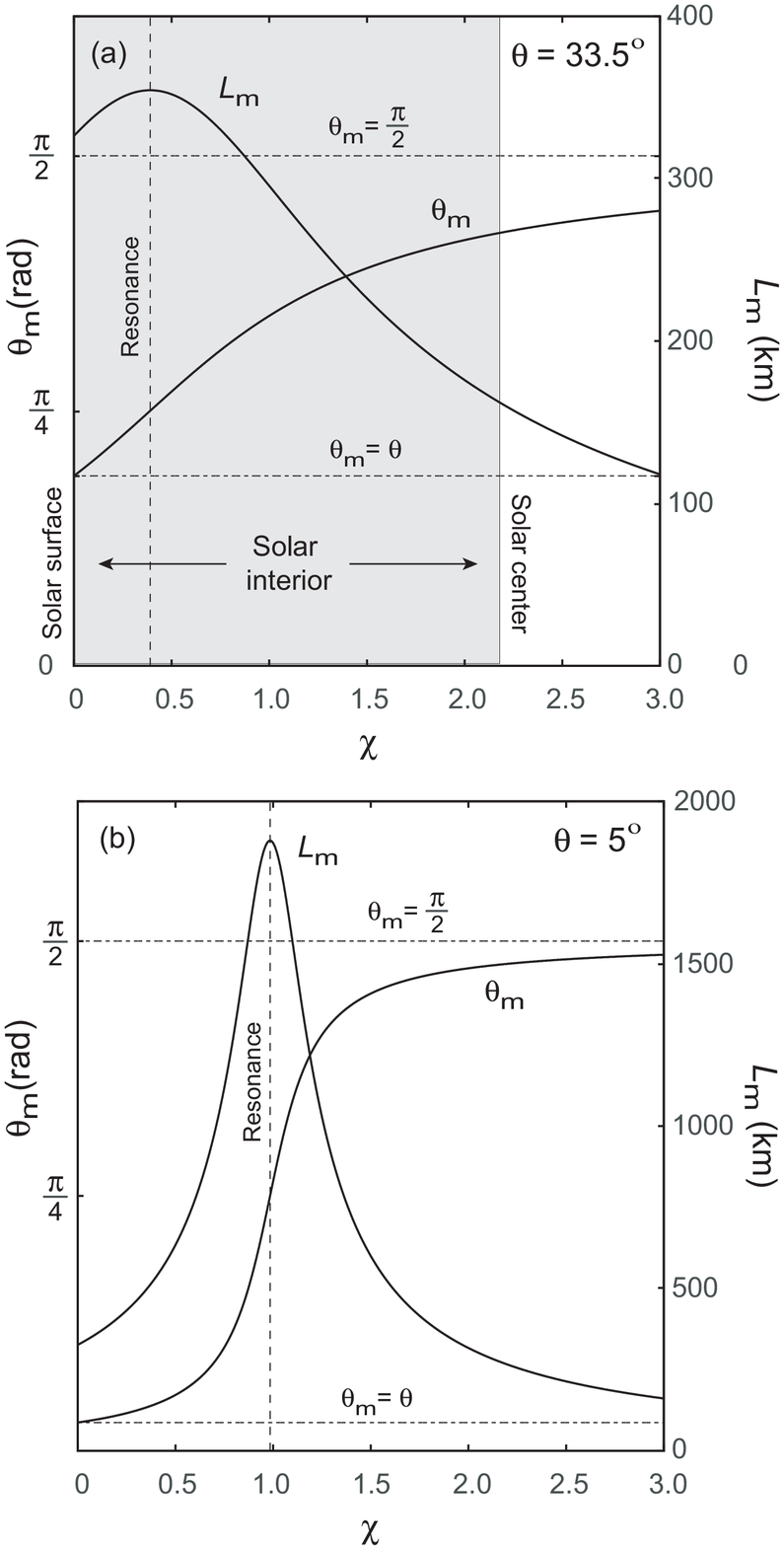}
%{figures/matterAngle_composite.eps}
{0pt}
{0pt}
{0.52}
{Matter mixing angle $\thetam$ as a function of the dimensionless coupling 
strength $\chi \equiv L/\ell\tsub m$ for vacuum mixing angles of (a)~$\thetav = 
33.5^\circ$ and (b)~$\thetav=5^\circ$.  Also shown is the matter oscillation length $L\tsub m$, which has a maximum at the  MSW resonance (see Section \ref{sna1.7.3}),  marked by the dashed vertical line. The oscillation 
length was computed from Eqs.\ (\ref{nosc19}) and (\ref{vosc8}) assuming $E = 
10\units{MeV}$ and $\Delta m^2 = 7.6\times 10^{-5} \units{eV}^2$. Case (a) is 
realistic for solar neutrinos and at the center of the Sun 
 $\chi \sim 2.13$. Hence the shaded 
region on the left side of (a) indicates the range of coupling strengths 
available to electron neutrinos in the interior of the Sun. }
The values of $\thetam$ and $L\tsub m$ will vary with the solar depth since they 
depend on the number density $n\tsub e$ through $\chi$.  From \eq{nosc18}, 
$\thetam \rightarrow \thetav$ as the electron density tends to zero, while in 
the opposite limit of very large electron density $\thetam \rightarrow 
\tfrac\pi2$. From \fig{matterAngle_composite}, the manner in which the two 
limits are approached depends on whether the vacuum mixing angle is large or 
small.  Case (a) corresponds to parameters valid for solar neutrinos.  At the 
solar center (where $\chi \sim 2.13$) the value of the matter mixing angle is $\thetam 
\sim 76^\circ$, compared with a vacuum mixing angle  $33.5^\circ$ at the solar 
surface. Conversely, for case (b) with $\thetav = 5^\circ$ at the solar surface, 
the matter mixing angle at a density corresponding to the solar center is 
$\thetav \sim 86^\circ$.

\subsection{\label{antineutrinoResonance} Resonance in the Antineutrino Channel}

If the condition $m_1 < m_2$ is not satisfied there is no MSW resonance for 
electron neutrinos in a 2-flavor model;  there is instead a corresponding 
resonance condition for the electron antineutrino $\bar \nu\tsub e$.   
Antineutrino oscillations could be important in environments such as 
core-collapse supernovae where all flavors of neutrinos and antineutrinos are 
found in copious amounts, but the Sun produces mostly neutrinos so consideration 
of oscillations for antineutrinos will be omitted from the present discussion.

\subsection{\label{sna1.7.4} Resonant Flavor Conversion}

If $m_1 < m_2$ and the electron density in the central region of the Sun where 
neutrinos are produced satisfies $n\tsub e > \resonanceDensity$, a solar 
neutrino will inevitably encounter the MSW resonance on its way out of the Sun. 
If the change in density is sufficiently slow that the additional phase mismatch 
between the $\nu\tsub e$ and $\nu_\mu$ components produced by the 
charged-current elastic scattering from electrons changes very slowly with 
density (the adiabatic condition discussed further below), the $\nu\tsub e$ flux 
produced in the core can be almost entirely converted to $\nu_\mu$ by the MSW 
resonance near the radius where the condition (\ref{nosc21}) is satisfied.  

MSW flavor conversion can be viewed as an {\em adiabatic 
quantum level crossing}, \cite{bet86,hax86} as illustrated schematically 
in \fig{flavorLevelCrossing},%
\singlefig
{flavorLevelCrossing}
{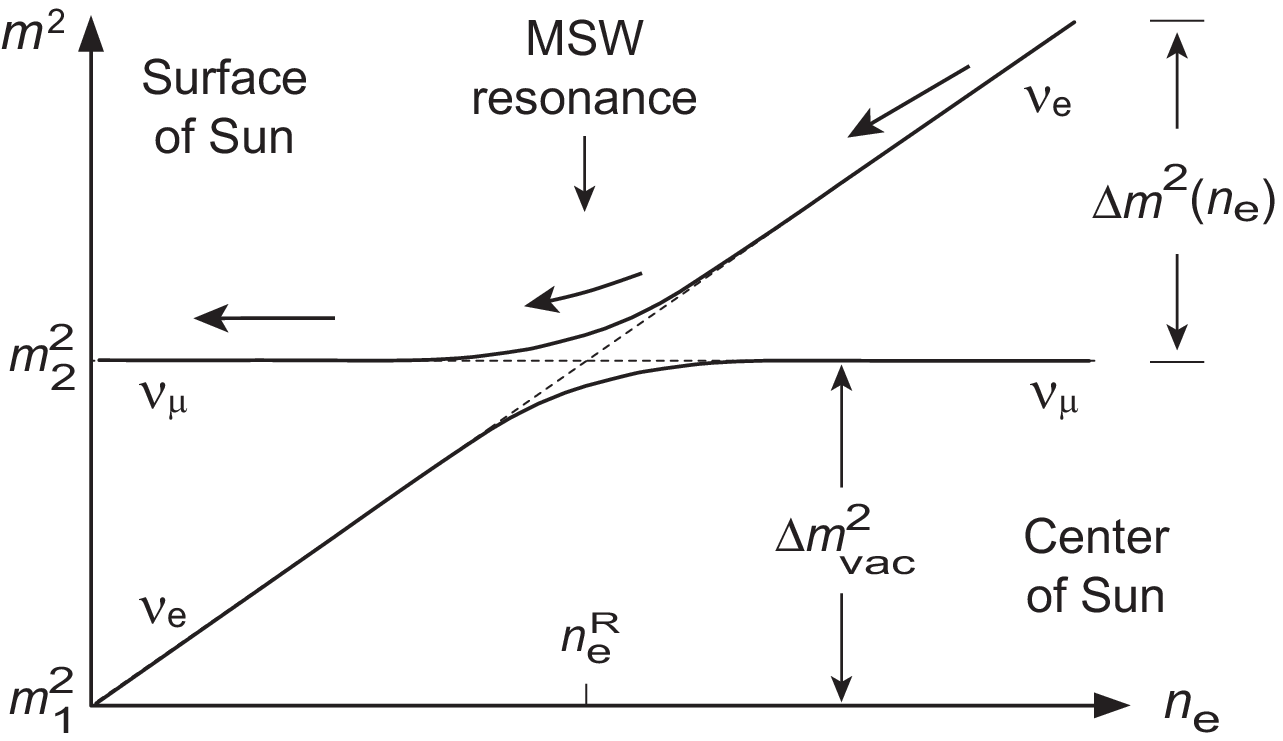}
{0pt}
{0pt}
{0.57}
{Schematic illustration of resonance flavor conversion as an adiabatic quantum 
level crossing. Specific examples for different parameter choices are shown in 
\fig{eigenvaluesMSW}.}
and more realistically in \fig{eigenvaluesMSW}.%
\singlefig
{eigenvaluesMSW}  
{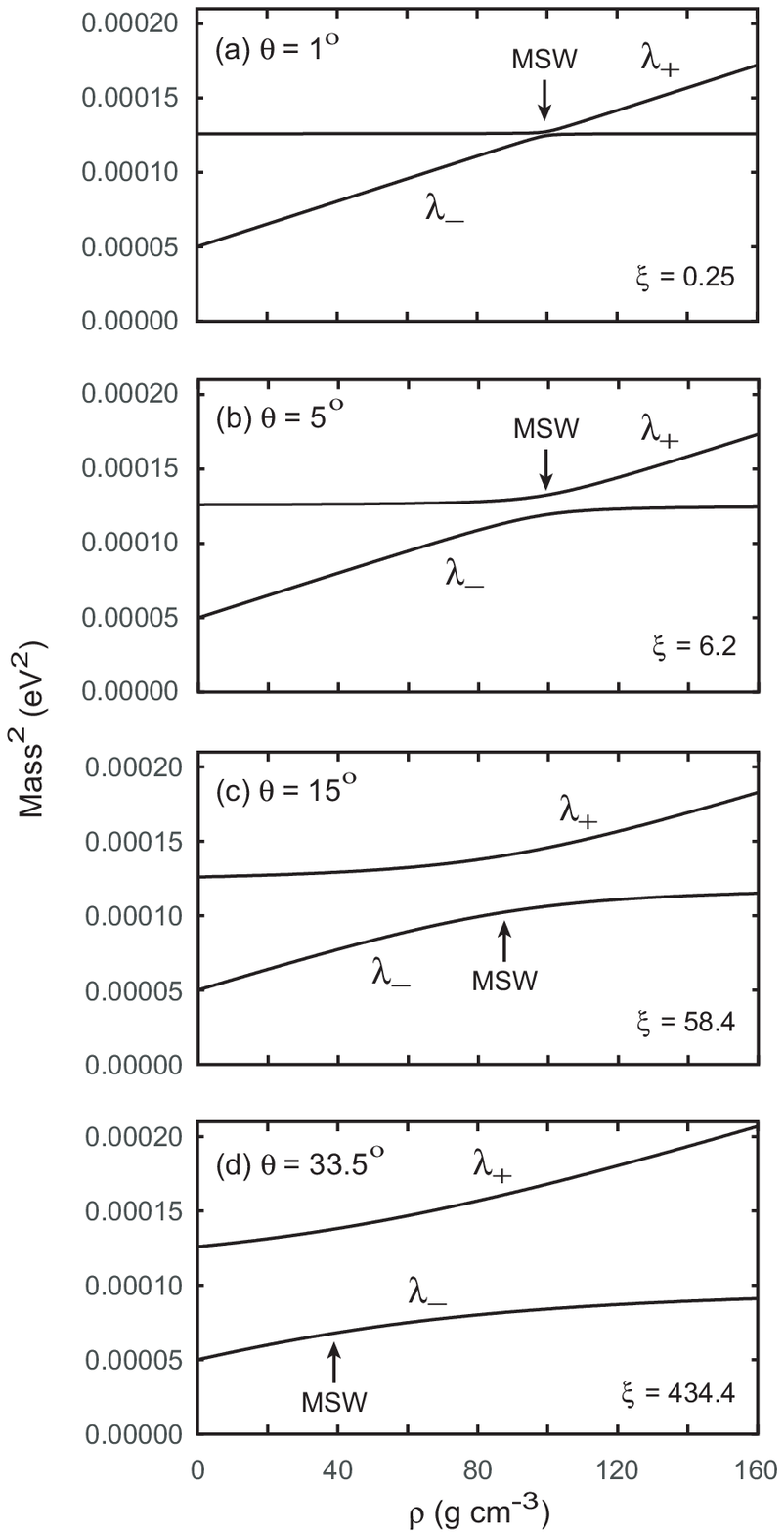} 
%{figures/eigenvaluesMSW_square2.eps} 
{0pt}   
{0pt}
{0.80}   
{Solutions $\lambda_\pm$ of the MSW eigenvalue problem  as a function of mass 
density according to \eq{matter1.4}.  Each case corresponds to the choices 
$\Delta m^2 = 7.6 \times 10^{-5} \units{eV}^2$ and $E=10\units{MeV}$, but to 
different values of the vacuum mixing angle $\thetav$. The individual neutrino 
masses are presently unknown but for purposes of illustration  $m_1^2 = 5\times 
10^{-5} \units{eV}^2$ has been assumed in vacuum, so that $m_2^2 = m_1^2 + 
\Delta m^2 = 1.26\times 10^{-4} \units{eV}^2$.  The critical density leading to 
the MSW resonance (corresponding to minimum splitting between the eigenvalues) 
and the value of the adiabaticity parameter $\xi=\delta r_{\rm\scriptscriptstyle 
R}/ L^{\rm\scriptscriptstyle R}_{\rm\scriptscriptstyle m}$ defined in 
\eq{adiabatic1.1} are indicated for each case. As will be discussed in 
Section \ref{sna1.8}, realistic solar conditions imply the 
highly adiabatic crossing exhibited in case (d). }
Solutions of the 
MSW eigenvalue problem illustrating this level crossing for various choices of 
the vacuum mixing angle are displayed in \fig{eigenvaluesMSW}. At zero 
density on the left side of these plots the eigenvalues $\lambda_\pm$ converge 
to the vacuum $m^2$ values, but for non-zero density the masses are altered by 
the interaction of the electron neutrino with the medium and the mixing of the 
solutions by the neutrino oscillation. (Compare with the unmixed case in 
\fig{effectiveNeutrinoMass}.) 

A neutrino produced  near the center of the Sun (high 
density on  right side) will be in a $\nu\tsub e$  
flavor eigenstate that coincides with the {\em higher-mass eigenstate,} since 
$V$ representing interaction with the medium increases the mass of the electron 
neutrino but not the muon neutrino. As the neutrino propagates out of the Sun 
(right to left in this diagram) the density decreases so $V$ and the effective 
mass of the neutrino decrease.  Conversely, the lower-mass eigenstate 
(primarily 
$\nu_\mu$ flavor) remains constant in mass as $n\tsub e$ decreases (no coupling 
to the charged current).  Thus the two levels cross at the critical density 
where the effect of $V$ exactly cancels the vacuum mass-squared difference 
$\Delta m^2_{\rm\scriptstyle vac} \equiv \Delta m^2(n\tsub e = 0)$ between the 
eigenstates, and the neutrino remains in the high-mass eigenstate and changes 
adiabatically into a $\nu_\mu$ flavor state by the time it exits the Sun (left 
side), because {\em in vacuum  the high-mass eigenstate approximately coincides 
with  $\nu_\mu$.} 

In summary, {\em if the crossing is adiabatic} the neutrino remains in the 
high-mass eigenstate in which it was created and follows the upper curved 
trajectory though the resonance in the level-crossing region, as indicated by 
the arrows.  It emerges  from the Sun in a  different flavor state than the one 
in which it was created in the core of the Sun because {\em the high-mass 
eigenstate is primarily $\nu\tsub e$ in the dense medium but is primarily 
$\nu_\mu$ in vacuum.}  Such adiabatic crossing of energy levels  is common in a 
variety of quantum systems and is often termed {\em avoided level crossing.} 

The number density of electrons in the Sun computed in the Standard Solar Model 
is illustrated in \fig{electronDensitySSM}, along with an exponential 
approximation that is rather good in the region of the Sun where the MSW effect 
is most important.   In \fig{criticalRadius},%
\singlefig
{criticalRadius} 
{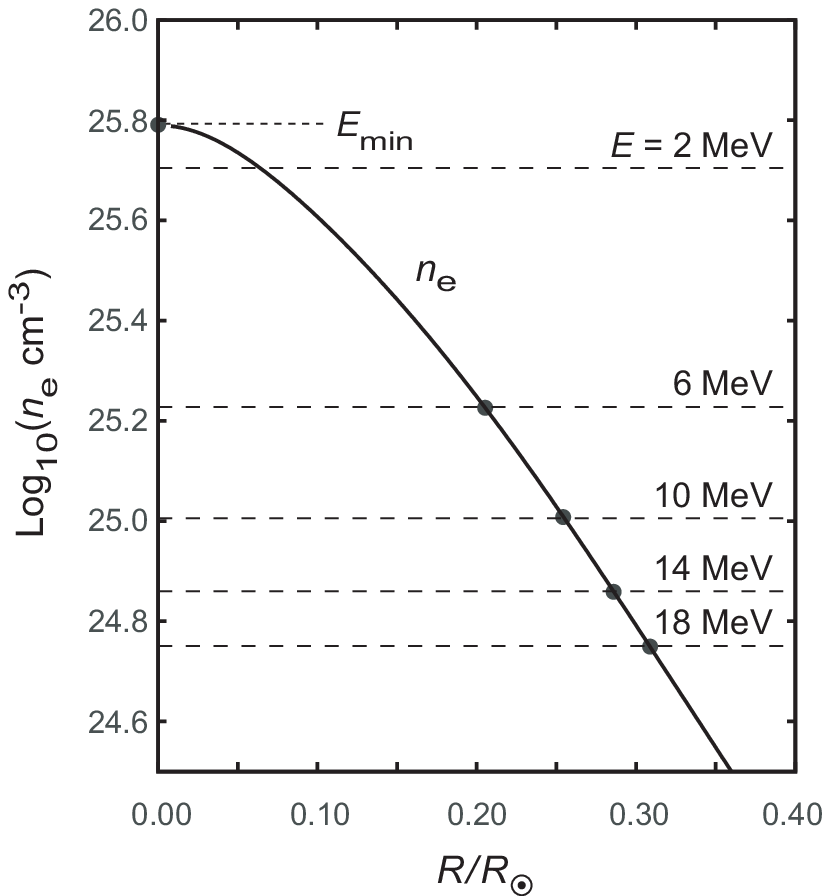}
{0pt}   
{0pt}        
{0.79}   
{Radius where the MSW critical density for a 2-flavor model is 
realized (dots at intersection of dashed lines with the curve for $n\tsub e$) 
for neutrinos of energies ranging from 2 to 18 MeV.  A vacuum mixing angle 
$\thetav = 35^\circ$ and $\Delta m^2c^4 = 7.5 \times 10^{-5} \units{eV}^2$ have 
been assumed. The minimum energy of an electron neutrino $E\tsub{min} \sim 1.6 
\units{MeV}$ that could be produced in the Sun and still encounter the MSW 
resonance is indicated.}
the 
approximate locations in the Sun where electron neutrinos of various energies 
would encounter the MSW resonance condition [determined by solving \eq{nosc21} 
for each choice of energy] are illustrated. Only neutrinos having an energy 
larger than some 
minimum energy $E\tsub{min}$  can experience the MSW resonance in the Sun 
because the neutrino must be produced at a density higher than the critical 
resonance density.  The conditions used to obtain \fig{criticalRadius}
give $E\tsub{min}\sim 1.6 \units{MeV}$. Thus, we may expect that the MSW 
effect is more efficient at converting the flavor of higher-energy neutrinos.  
That flavor conversion is observed to be preferentially suppressed for 
lower-energy solar neutrinos will be an important piece of 
evidence favoring the MSW mechanism over vacuum oscillations for the observed 
flavor conversion of solar neutrinos.

\subsection{\label{varyingDensityMatter} Neutrino Propagation in Matter with 
Varying Density}

Consider now realistic neutrino propagation in the Sun, where 
a neutrino produced in the core will encounter steadily decreasing density as it 
travels toward the solar surface (see \fig{solarLayers}). The 
neutrino flavor evolution will be governed by the analog of the coupled 
differential equations for vacuum propagation, but with $U \rightarrow U\tsub 
m(t)$ since the flavor--mass basis transformation now depends on time.  From 
\eq{nmm1.6} with this replacement
$$
 i\,\frac{d}{\diffelement t} \left[ U\tsub m(t) 
\onematrix{\nu_1(t)}{\nu_2(t)} \right]
= \frac{1}{2E} \, U\tsub m(t) 
\twomatrix
{m_1^2}
{0}
{0}
{m_2^2}
\onematrix{\nu_1(t)}{\nu_2(t)} ,
%\label{propm1.1}
$$
where both the wavefunctions and the transformation matrix $U\tsub m$ are 
indicated explicitly to depend on the time.  Taking the time derivative of the 
product in brackets on the left side and multiplying the equation from the left 
by $U_{\rm\scriptstyle m}^\dagger$ gives 
\begin{align}
  i\frac{d}{\diffelement t} 
\onematrix{\nu_1}{\nu_2}
&=
\left[
\twomatrix
{m_1^2/2E}
{0}
{0}
{m_2^2/2E}
-i\,U^\dagger_{\rm\scriptstyle m} \,\frac{d}{\diffelement t} \,U\tsub m
\right]
\onematrix{\nu_1}{\nu_2}
\nonumber
\\
&=
\twomatrix
{m_1^2/2E}
{-i\dotthetam}
{i\dotthetam}
{m_2^2/2E}
\onematrix{\nu_1}{\nu_2}
\nonumber
\\
&=
\twomatrix
{-\Delta m^2/4E}
{-i\,\dotthetam}
{i\,\dotthetam}
{\Delta m^2/4E}
\onematrix{\nu_1}{\nu_2} ,
\label{propm1.2}
\end{align}
where \eq{matter1.5b} was used,  $\dot\theta\tsub m \equiv d\thetam/dt$, and in 
the last step the constant $ (m_1^2 +m_2^2)/4E$ times the unit matrix has been 
subtracted from the matrix on the right side (which does not affect 
observables).

Our earlier statement that mass eigenstates at some density in the Sun generally 
will not be eigenstates at a different density may now be quantified. If the 
mass matrix in \eq{propm1.2} were diagonal the neutrino would remain in its 
original mass eigenstate as it traveled through regions of varying density, so 
it is the off-diagonal terms proportional to $\dot\theta\tsub m = d\thetam/dt$ 
that alter the mass eigenstates as the neutrino propagates.  Generally then, 
\eq{propm1.2} must be solved numerically.  However, if the off-diagonal terms 
are small relative to the diagonal terms, the mass matrix $M$ may be 
approximated by dropping the off-diagonal terms
\begin{equation}
 M = \twomatrix
{-\Delta m^2/4E}
{-i\,\dotthetam}
{i\,\dotthetam}
{\Delta m^2/4E}
\simeq
\frac{\Delta m^2}{4E}
 \twomatrix
{-1}
{0}
{0}
{1} ,
\label{propm1.3}
\end{equation}
which affords the possibility of an analytical solution for neutrino flavor 
conversion in the Sun.  This is called the {\em adiabatic approximation}, and 
corresponds physically to the assumption that the matter mixing angle $\thetam$ 
changes only slowly over a characteristic time for motion of the neutrino. A 
neutrino travels at near light speed so $r\sim ct$ and the adiabatic condition 
also may  be interpreted as a limit on the spatial gradient of $\thetam$.  From 
\fig{matterAngleLength}(a) and \fig{matterAngle_composite}, the most rapid 
change in $\thetam$ and thus the largest contribution of the off-diagonal terms 
occurs near the MSW resonance (corresponding to $\thetam = \frac\pi4$). Let us 
now use these observations to quantify the conditions appropriate for the 
adiabatic approximation.

\subsection{\label{sna1.7.4b} The Adiabatic Criterion}

The adiabatic condition for resonant flavor conversion can be expressed as a 
requirement that the spatial width of the resonance layer $\delta 
r_{\rm\scriptscriptstyle R}$ (which is defined by the radial distance over which the 
resonance condition is approximately fulfilled) be much greater than the 
oscillation wavelength in matter evaluated at the resonance, 
$L^{\rm\scriptscriptstyle R}_{\rm\scriptscriptstyle m}$.  
This can be characterized by introducing an {\em adiabaticity parameter} $\xi$ through the definition [see \eq{nosc23}]
\begin{equation}
\xi \equiv \frac{\delta r_{\rm\scriptscriptstyle R}}{L^{\rm\scriptscriptstyle 
R}_{\rm\scriptscriptstyle m}}
\quad
\delta r_{\rm\scriptscriptstyle R} =
\frac{\resonanceDensity}{(\diffelement n\tsub 
e/\diffelement r)_{\rm\scriptscriptstyle R}}
\tan 2\thetav
\quad 
L^{\rm\scriptscriptstyle R}_{\rm\scriptscriptstyle m} = \frac{L}{\sin 2\thetav} 
,
\label{adiabatic1.1}
\end{equation}
 where an index 
 {\small R} labels quantities that are evaluated at the resonance, $L$ is the 
vacuum oscillation length, and $\thetav$ is the vacuum oscillation angle. 
\cite{bet86,smir2003}  The adiabatic condition corresponds then to 
the requirement that $\xi \muchgreater 1$, which implies physically that if 
many flavor oscillation lengths (in matter) fit within the resonance layer the 
adiabatic approximation of Eq.\ (\ref{propm1.3}) is valid.

Values of  $\xi$ computed from \eq{adiabatic1.1} are indicated in 
\fig{eigenvaluesMSW} for several choices of vacuum mixing angle $\thetav$.  In 
general  very sharp level crossings  as in \fig{eigenvaluesMSW}(a) are 
non-adiabatic, while avoided level crossings as in \fig{eigenvaluesMSW}(d) are 
highly adiabatic. In the limit of no mixing ($\thetav=0$) the levels cross with 
no interaction, as illustrated earlier in \fig{effectiveNeutrinoMass}. From 
\fig{eigenvaluesMSW} one sees that the MSW resonance can occur under 
approximately 
adiabatic conditions, even for relatively small values of the vacuum mixing 
angle [for example, case (b)]. As will be shown in Section \ref{sna1.8}, solar 
conditions correspond to the highly-avoided level crossing in 
\fig{eigenvaluesMSW}(d), for which $\delta r_{\rm\scriptscriptstyle R} 
\muchgreater L^{\rm\scriptscriptstyle R}_{\rm\scriptscriptstyle m}$.  Thus 
the MSW resonance is expected to be encountered  adiabatically in the Sun, 
which 
optimizes the chance of resonant flavor conversion.

\subsection{\label{sna1.7.4c} MSW Neutrino Flavor Conversion}

Generally neutrino flavor conversion in the Sun must be obtained by integrating 
\eq{propm1.2} numerically because of the solar density gradient, but it has just been
argued that the adiabatic approximation (\ref{propm1.3}) should be very well 
fulfilled for the Sun. Therefore, let's now solve for flavor conversion of solar 
neutrinos 
by the MSW mechanism, assuming the validity of the adiabatic approximation. 

\subsubsection{Solar Flavor Conversion in Adiabatic Approximation}

The adiabatic conversion of neutrino flavor in the Sun is illustrated in 
\fig{adiabaticFlavors}.%
\doublefig
{adiabaticFlavors}   
{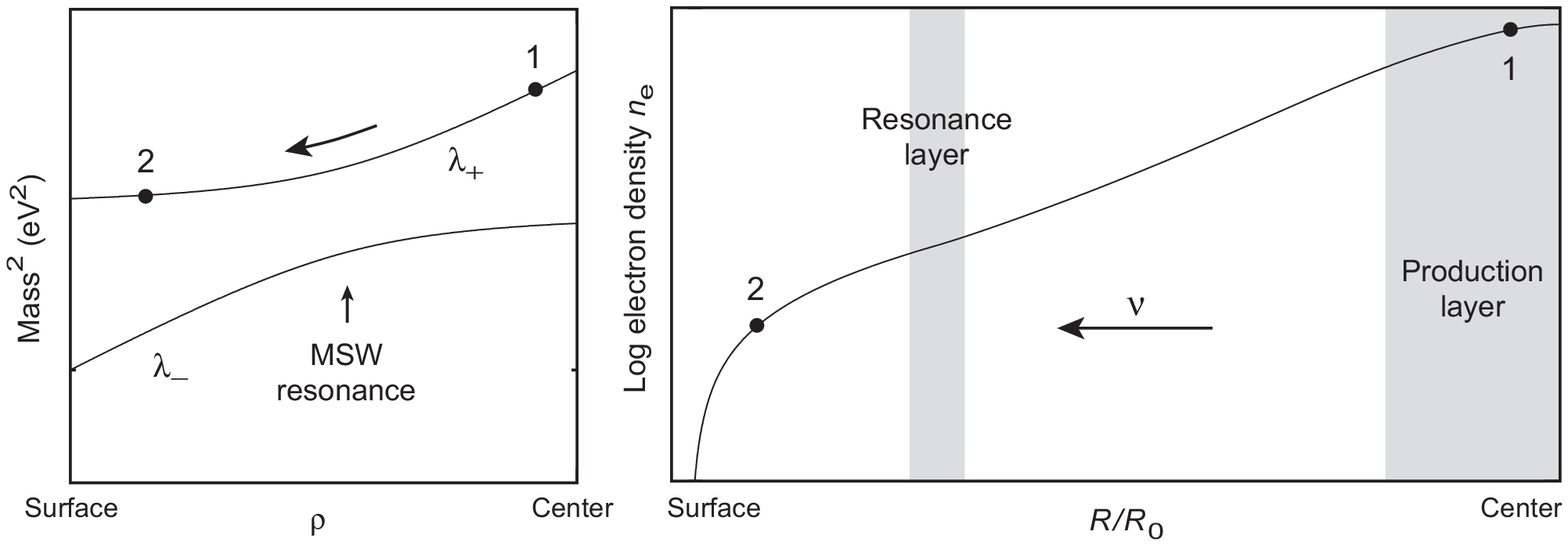} 
{0pt}   
{0pt}       
{0.73}   
{Adiabatic flavor conversion by the MSW mechanism in the Sun.  An electron 
neutrino is produced at Point 1 where the density lies above that of the MSW 
resonance and propagates radially outward to Point 2 where the density 
lies below that of the resonance. The width of the resonance layer is assumed 
to be much larger than the matter oscillation length in the resonance layer, 
justifying the adiabatic approximation of \eq{propm1.3}. Widths of resonance 
and production layers are not meant to be to scale in this figure.
}
An electron neutrino is produced at Point 1 near the center of the Sun and 
propagates radially outward to Point 2. Detection is assumed to average over 
many oscillation lengths so that the interference terms are washed out and our
concern is with the classical (time-averaged) probability, as described in 
Section \ref{timeAveragedVac}. The probability to be detected at Point 2 in the 
$\ket{\nu\tsub e}$ flavor eigenstate is then given by \cite{kuo1989}
\begin{align*}
\peec
&=
(1\ \ \ 0)
\twomatrix
{\cos^2\thetam(2)}
{\sin^2\thetam(2)}
{\sin^2\thetam(2)}
{\cos^2\thetam(2}
\nonumber\\
&\times
\unittwo
\twomatrix
{\cos^2\thetam(1)}
{\sin^2\thetam(1)}
{\sin^2\thetam(1)}
{\cos^2\thetam(1)}
\onematrix
{1}
{0} ,
 %\label{adiabatic1.2}
\end{align*}
where $\thetam(i) \equiv \thetam(t_i)$ and the row vector $(1\ 0)$ and 
corresponding column vector  denote a pure $\nu\tsub e$ flavor state.
Evaluating the matrix products and using standard trigonometric identities,
\begin{equation}
\peec
=
\tfrac12 \left[ 1+ \cos 2\thetam(t_1) \cos 
2\thetam(t_2) \right] .
\label{adiabatic1.3}
\end{equation}
This result is valid (if the adiabatic condition is satisfied) for Point 2 
anywhere outside Point 1,
but in the specific case that Point 2 lies at the solar surface $\thetam(t_2) 
\rightarrow \thetav$ and the classical probability to detect the neutrino as an 
electron neutrino when it exits the Sun is
\begin{equation}
\peec
=
\tfrac12 \left[ 1+ \cos 2\thetav \cos 
2\theta^{\scriptscriptstyle 0}_{\rm\scriptscriptstyle m} \right] ,
% \qquad
% \units{(at\ the\ solar\ surface)} ,
 \label{adiabatic1.4}
\end{equation}
where $\thetav$ is the vacuum mixing angle and $ \theta^{\scriptscriptstyle 
0}_{\rm\scriptscriptstyle m}\equiv \thetam(t_1)$ is the matter mixing angle at 
the point of neutrino production.

\subsubsection{Dependence Only on the Mixing Angle}

Equation (\ref{adiabatic1.4}) has a simple physical interpretation. The mass 
matrix for a neutrino propagating down the solar density gradient is diagonal by 
virtue of the adiabatic assumption (\ref{propm1.3}), so a neutrino produced in 
the $\lambda_+$ eigenstate remains in that mass eigenstate until it reaches the 
solar surface, with the flavor conversion resulting only from the change of 
mixing angle between the production point and the surface.  Thus, in adiabatic 
approximation the classical probability $\peec$ depends only on the mixing 
angles at the point of production and point of detection, and is {\em 
independent of the details of neutrino propagation.} This is reminiscent of the 
results obtained in Section \ref{timeAveragedVac}  for the classical average of vacuum 
oscillations; indeed,  \eq{adiabatic1.4} is equivalent to \eq{flavorProbsAVG} in 
the limit that $\thetam \rightarrow \thetav$.
MSW flavor conversion in adiabatic 
approximation is illustrated for four different values of the vacuum mixing 
angle $\thetav$ in \fig{flavorVsR}.
\doublefig
{flavorVsR}
{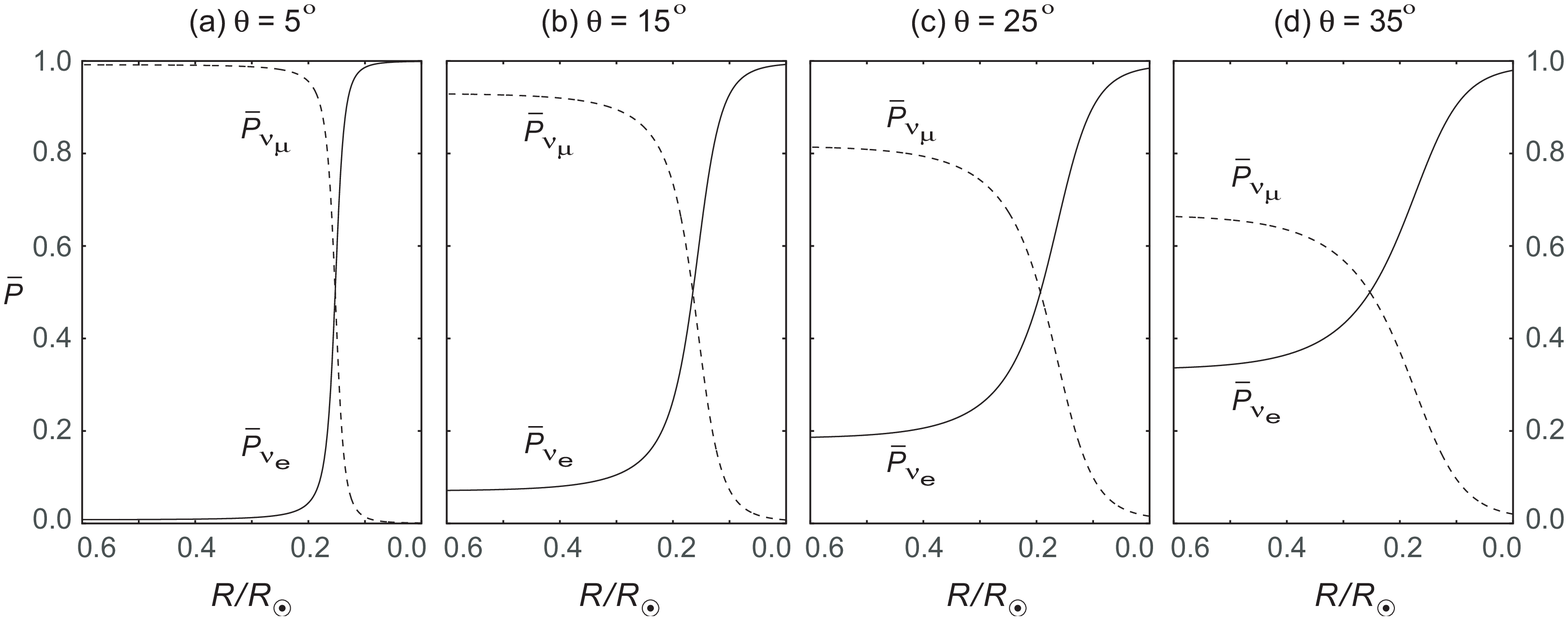} 
{0pt}   
{0pt}       
{0.45}   
{MSW flavor conversion  versus fraction of solar radius for four values of the 
vacuum mixing angle $\thetav$ in a 2-flavor model.    All calculations are 
classical averages (Section \ref{timeAveragedVac}) over local oscillations in 
adiabatic approximation using \eq{adiabatic1.4} with $\Delta m^2 = 
7.6\times 10^{-5} \punits{eV}{2}$ and $E = 10 \units{MeV}$. The exponential 
density approximation shown in \fig{electronDensitySSM} was used and  neutrinos 
were assumed to be produced in a $\nu\tsub e$ flavor state at the center (right 
side of diagram at $R/R_\odot = 0$). Solid curves show the classical 
electron-neutrino probability $\bar P_{\nu_{\scriptscriptstyle\rm e}} \equiv 
\peec$ and  dashed curves show the corresponding classical muon-neutrino 
probability $\bar P_{\nu_{\mu}} \equiv \pemuc$.  In these figures the MSW 
resonance occurs at the radius corresponding to the 
intersection of the solid and dashed curves.}
  Figure \ref{flavorVsR}(d) 
approximates the situation expected for the Sun.

\subsubsection{Resonant Conversion for Large or Small $\bm \thetav$}

As shown in \fig{matterAngleLength}(a),  the matter mixing angle $\thetam$ 
approaches  $\frac\pi2$ near the center of the Sun and becomes equal to 
$\thetav$ at the surface of the Sun.  Hence  neutrinos are produced in a flavor eigenstate 
that is an almost pure mass eigenstate, but they evolve to a flavor mixture 
characterized by the vacuum mixing angle $\thetav$ by the time they exit the 
Sun. The most rapid flavor conversion occurs around the MSW resonance where the 
$P_{\nu_{\scriptscriptstyle\rm e}}$ and $P_{\nu_{\mu}}$ curves intersect. For 
smaller vacuum mixing angle $\thetav$ almost complete flavor conversion occurs in the resonance, 
while for the large mixing angle case of \fig{flavorVsR}(d) about $\frac23$ of  
10-MeV $\nu\tsub e$ produced in the core will undergo flavor conversion before 
leaving the Sun.

\subsubsection{Energy Dependence of Flavor Conversion}

Figure \ref{criticalRadius} suggests that flavor conversion has a significant energy 
dependence.  For example, repeating the calculation of \fig{flavorVsR}(d) for a 
range of neutrino energies $E$ gives the electron neutrino survival 
probabilities displayed in Table \ref{flavorEnergyDependence},
%
%
% The entries in this table were calculated using flavorVsR.gnu
%
\begin{table*}[t]
\centering
\caption{Energy dependence of solar neutrino flavor conversion for 
a vacuum mixing angle $\thetav=35^\circ$}
\begin{ruledtabular}
\begin{tabular}{ccccccc}
$E$ (MeV) & 14 &  10 & 6 & 2 & 1 & 0.70
\\
\hline
$P_{\nu_{\scriptscriptstyle\rm e}}$ (surface)& 0.33 & 0.33 & 0.34 & 0.40 & 0.47 
& 0.50
\\
$R^{\rm\scriptscriptstyle R}/R_\odot$ & 0.28 & 0.25 & 0.20 & 0.10 & 0.03 & 0.0
\\
%\hline
\end{tabular}
\end{ruledtabular}
\label{flavorEnergyDependence}
\end{table*}
along with the fractional solar radius $R^{\rm\scriptscriptstyle R}/R_\odot$ 
where the MSW resonance occurs for that energy. From the  
spectrum in \fig{solarNeutrinoSpectrum} and the experimental neutrino 
anomalies of Table \ref{snusExp}, one sees an overall 
suppression of expected electron neutrino probabilities to the 30\%--50\% range 
by the MSW effect, with the lower end of this range associated with 
higher-energy neutrinos.  These results suggest a possible resolution of the 
solar neutrino problem that will be elaborated further in Section \ref{sna1.8}.

\section{\label {sna1.8} Resolving the Solar Neutrino Problem}

The two-flavor neutrino oscillation formalism described above may now be used in 
conjunction with a series of key neutrino observations by Super Kamiokande, the 
Sudbury Neutrino Observatory (SNO), and KamLAND to resolve the solar neutrino 
problem. As will now be summarized, this analysis indicates that electron 
neutrinos are being converted to other flavors by neutrino 
oscillations, that  the solar neutrino ``deficit''  disappears if all flavors of neutrinos coming from the Sun are detected, 
and that the favored oscillation scenario is MSW resonance conversion in the Sun 
for a large vacuum mixing angle solution. 

\subsection{\label{sna1.8.1} Super Kamiokande Observation of Flavor Oscillation}

Cosmic rays striking the atmosphere generate showers of mesons that 
decay to muons, electrons, positrons, and neutrinos. The Super Kamiokande 
(Super K) detector in Japan was used to observe neutrinos produced in 
these atmospheric cosmic ray showers.  These measurements found  that the 
ratio of muon neutrinos and antineutrinos to electron neutrinos and 
antineutrinos was less than $\sim 64\%$ of the value expected from the 
Standard Model, 
and the results were interpreted as an indication that the muon neutrino was 
undergoing oscillations with another flavor neutrino that was not the electron 
neutrino. \cite{fuk98}  This was the first conclusive evidence of neutrino 
oscillations and thus of finite neutrino mass.

\subsection{\label{sna1.8.2} SNO Observation of Neutral Current Interactions}

The Super Kamiokande atmospheric neutrino results cited above indicate the existence of 
neutrino oscillations and thus of physics beyond the Standard Model.  The 
neutrino oscillations discovered by Super Kamiokande
are not directly applicable to the solar neutrino problem 
because they do not appear to involve electron neutrinos.  However, a modified 
water \cerenkov\ detector operating in Canada has yielded information about neutrino 
oscillations that does have implications for  the solar neutrino problem.

\subsubsection{SNO and Heavy Water}

The Sudbury Neutrino Observatory (SNO) differed from Super Kamiokande in that it 
contained 
heavy water (water enriched in deuterium, \isotope2H) at its core.  In regular 
water solar neutrinos can signal their presence only by elastic scattering from 
electrons, which typically requires  about 5--7 MeV of energy to produce \cerenkov\ light for 
reliable detection.  However, the deuterium (d) contained in the heavy water  can undergo a 
breakup reaction when struck by a neutrino  through the {\em 
weak neutral current,} where any flavor neutrino can initiate the reaction
\begin{equation}
  \nu + d \rightarrow \nu + p + n \qquad \units{(Neutral current) ,} 
   \label{nosc25}
\end{equation}
and the {\em charged weak current reaction}
\begin{equation}
  \nu\tsub e + d \rightarrow e^- + p + p \qquad \units{(Charged current)},
   \label{nosc26}
\end{equation}
which can be initiated only by electron neutrinos.
These reactions have much larger cross sections than elastic 
neutrino--electron scattering, and the energy threshold can be 
lowered to 2.2 MeV, the binding energy of the deuteron. Crucially, the 
neutral-current reaction (\ref{nosc25}) is flavor blind, which gave SNO the 
ability to see the {\em total neutrino flux of all flavors} coming from the Sun.

\subsubsection{Observation of Flavor Conversion for Solar Neutrinos}

The SNO results are summarized concisely in Table \ref{SNOresultsTable} and 
\fig{SNOresults}(a).%
\begin{table*}[t]
\centering
\caption{Comparison of SNO results and Standard Solar Model (SSM) 
for solar neutrino fluxes. \cite{ahm02a,ahm02b}  
Fluxes in units of $10^6 \ {\rm cm}^2\,{\rm s}^{-1}.$}
\begin{ruledtabular}
\begin{tabular}{ccccc}
SSM $\nu\tsub e$ flux & SNO $\nu\tsub e$ flux &  SNO $\nu\tsub e$/SSM &
SNO all flavors & SNO all/ SSM
\\
\hline
$5.05 \pm 0.91$ & $1.76 \pm 0.11$ & 0.348 & $5.09 \pm 0.62$ & 1.01
\\
\end{tabular}
\end{ruledtabular}
\label{SNOresultsTable}
\end{table*}
\doublefig
{SNOresults}           % Latex Cross-Ref Label
{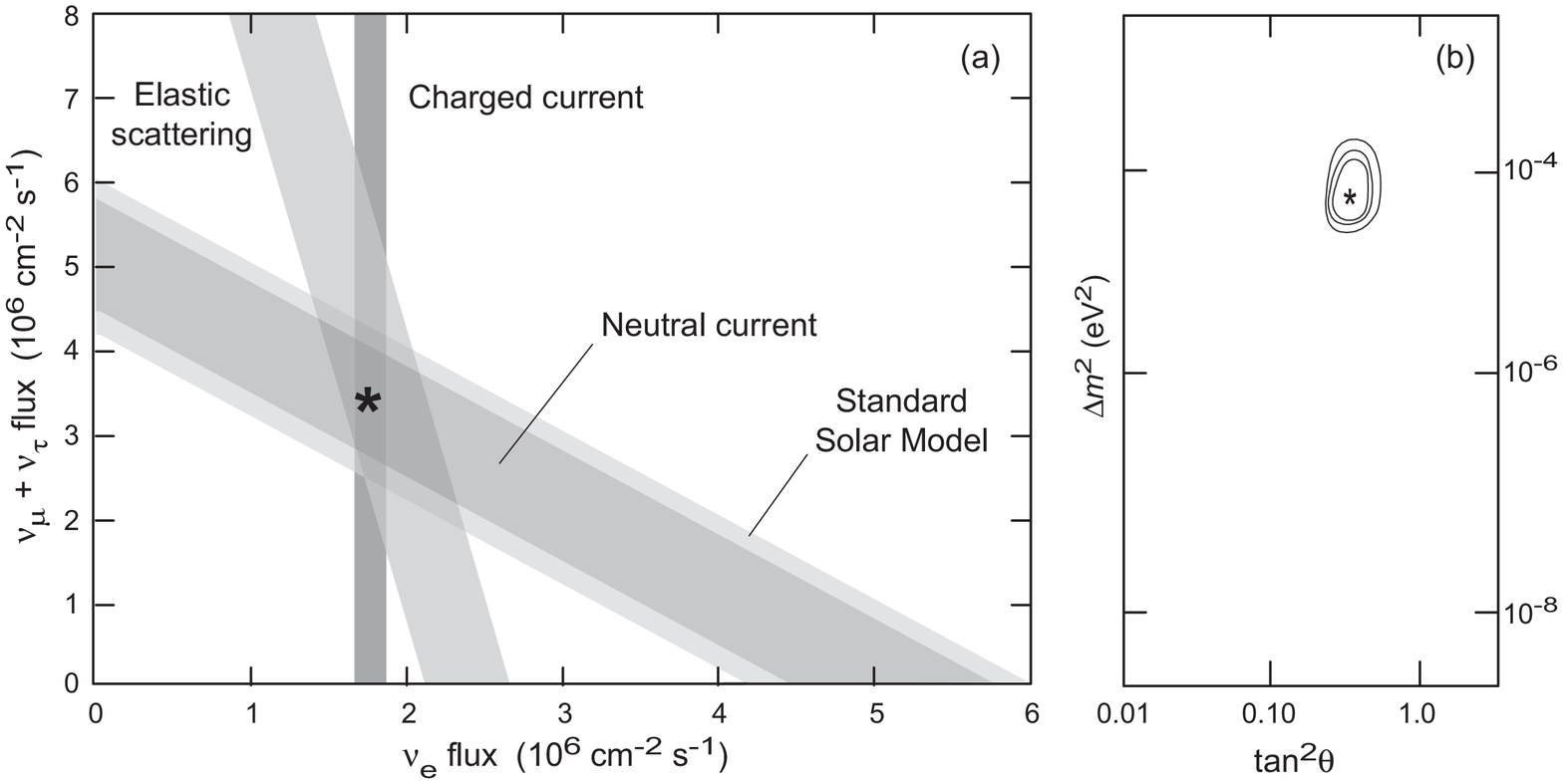}
{0pt}             % Extra space above, with dimensions.tex
{0pt}             % Extra space below, with dimensions.tex
{0.60}            % Scale factor, 1 is full
{(a)~Flux of solar neutrinos from \isotope8B  detected for various 
flavors by SNO. \cite{ahm02a} The band widths represent one 
standard deviation.  Note that the bands intersect at the point indicated by the 
star, which implies that about $\frac23$ of the Sun's \isotope8B 
neutrinos have changed flavor between being produced in the core of the Sun and 
being detected on Earth. The Standard Solar Model band is the prediction for the 
\isotope8B flux, irrespective of flavor changes. Note that it tracks the 
neutral current band, which represents detection of all flavors of neutrino 
coming from the Sun. (b)~2-flavor neutrino oscillation parameters determined by
SNO \cite{ahm02b}. The 99\%, 95\% and 90\% confidence-level contours are 
shown, with the star at the most likely value. The best fit corresponds to the 
large-angle solution given in Eq.~(\ref{nosc27}). }
The SNO observations confirmed results from the pioneering solar neutrino 
experiments summarized in Table \ref{snusExp}: a strong suppression of the 
electron neutrino flux is observed relative to that expected in the Standard 
Solar Model. \cite{ahm02a} However, by analyzing the flavor-blind weak 
neutral current events, the total flux of all neutrinos  was found to be almost 
exactly that expected from the Standard Solar Model.  Table \ref{SNOresults} 
summarizes: although the $\nu\tsub e$ flux is only 35\% of that expected without 
oscillations, the flux summed over all flavors is 100\% of that predicted by the 
Standard Solar Model for electron neutrino emission, within the experimental 
uncertainty.

The SNO case  was  strengthened by analysis of neutrino--electron elastic 
scattering data 
combined 
with data from the charged-current reaction (\ref{nosc26}).  This also allows an 
estimate of the total neutrino flux of all flavors: in elastic scattering from 
electrons, both charged and neutral currents contribute for electron neutrinos 
but only neutral currents do so  for other flavors. Figure \ref{SNOresults}(a) 
illustrates the flux of neutrinos from the \isotope8B reaction, based on SNO 
results.  The best fit indicates that $\frac23$ of the Sun's electron neutrinos 
have changed flavor by the time they reach  Earth.

\subsubsection{SNO Flavor-Mixing Solution}

Figure \ref{SNOresults}(b)
shows the confidence-level contours for  SNO data, 
which suggest that the solar neutrino problem is solved by $\nu\tsub 
e$--$\nu_\mu$ flavor oscillations with
\begin{equation}
  \Delta m^2 = 6.5_{-2.3}^{+4.4} \times 10^{-5} \punits{eV}{2}  \qquad 
\thetav = 33.9^{+2.4\, ^\circ}_{-2.2\, ^\circ} .
   \label{nosc27}
\end{equation}
This  {\em large-mixing-angle solution} means that  $\nu\tsub e$ 
is almost an equal superposition of two mass eigenstates, separated by at most 
a few hundredths of an eV.

\subsection{\label{sna1.8.3} KamLAND Constraints on Mixing Angles} 

KamLAND  used phototubes to monitor 
a large container of liquid scintillator, looking specifically for electron 
antineutrinos produced during nuclear power generation in a set of nearby 
Japanese 
and Korean reactors. Antineutrinos are detected from the inverse 
$\beta$-decay in the 
scintillator:
$
   \bar\nu\tsub e + p \rightarrow e^+ + n.
$
From the power levels in the reactors, the 
expected antineutrino flux at KamLAND could
be modeled accurately.  The experiment detected a shortfall of 
antineutrinos relative to the expected number and this could be accounted 
for assuming (anti)neutrino oscillations with a large-angle
solution having \cite{egu03,kam2008}
\begin{equation}
   \Delta m^2 = 7.58 ^{+0.14}_{-0.13}
%\units{(stat)}^{+0.15}_{-0.15} \units{(sys)}
\times 10^{-5} \units{eV}^2 \qquad 
\tan^2\thetav = 0.56^{+0.10}_{-0.07}   
\label{nosc29}
\end{equation}
(only statistical uncertainties are indicated), which corresponds to 
$\thetav \sim 36.8^\circ$ for the vacuum mixing angle.%

The oscillation properties of neutrinos and antineutrinos of the same generation 
are expected to be equivalent by CPT symmetry, where C is charge conjugation, P is parity, and T is time reversal.  Thus the large-angle KamLAND 
solution for electron antineutrinos may be interpreted as corroboration of the 
large-angle solution found for solar neutrinos.
Combining the solar neutrino and KamLAND results leads to a solution
\cite{kam2008}
\begin{equation}
   \Delta m^2 = 7.59 \pm 0.21
\times 10^{-5} \units{eV}^2 \quad 
\tan^2\thetav = 0.47^{+0.06}_{-0.05}   ,
\label{nosc29bb}
\end{equation}
implying a vacuum mixing angle $\thetav \sim 34.4^\circ$. The solar neutrino 
problem is resolved by neutrino oscillations for which the vacuum mixing angle 
is {\em large} (recall that $\thetav$ is defined so that its maximum value is 
$45^\circ$).

\subsection{\label{mswLargeAngle} Large Mixing Angles and the MSW Mechanism}

The large mixing angle solutions found by SNO and KamLAND indicate that the 
vacuum oscillations of solar neutrinos are of secondary importance to the MSW 
matter oscillations in the body of the Sun itself in reducing the electron neutrino flux.  The large-angle solutions 
imply vacuum oscillation lengths of a few hundred kilometers, so the classical 
average (\ref{flavorProbsAVG}) applies and for $\thetav \sim 34^\circ$ the 
reduction in $\nu\tsub e$ flux from averaging over vacuum oscillations is by 
about $\sim57\%$. Thus for vacuum oscillations the suppression of the electron 
neutrino flux detected on Earth would be by a factor of less than two, but the 
Davis chlorine experiment indicates a suppression by a factor of three. flavor 
conversion more severe than is possible from vacuum oscillations  seems 
required, and this can be explained by the MSW resonance, as has been 
illustrated in \fig{flavorVsR}. Indeed, \fig{flavorVsR}(d) indicates that for 10 
MeV neutrinos and parameters consistent with \eq{nosc27}, the MSW resonance 
gives a $\nu\tsub e$ suppression  by a factor of three.

Furthermore, vacuum oscillation lengths for the large-angle solutions are much 
less than the Earth--Sun distance, which would largely wash out any energy 
dependence of the electron neutrino shortfall. Since the observations indicate 
that such an energy dependence exists (see Table \ref{snusExp}), and the MSW 
effect implies such an energy dependence (see \fig{electronDensitySSM} and 
Table \ref{flavorEnergyDependence}), the MSW resonance is  implicated as the 
primary source of the neutrino flavor conversion responsible for the ``solar 
neutrino problem''.  That is, the MSW resonance converts from $\frac12$ to 
$\frac23$ (depending on the neutrino energy) of electron neutrinos into other 
flavors within the body of the Sun, and these populations are then only 
somewhat modified by the vacuum oscillations before the solar neutrinos reach 
detectors on Earth.

\subsection{\label{sna1.8.4tale} A Tale of Large and Small Mixing Angles}

Initial theoretical prejudice favored a small 
vacuum mixing angle. The MSW effect attracted 
initial attention because it was hoped that it might explain how a {\em small 
mixing angle} could account for the solar neutrino deficit, 
since  \fig{flavorVsR} indicates that the MSW resonance can generate almost 
complete flavor conversion even for small vacuum mixing angles.  However, data 
now indicate that the MSW resonance is indeed the solution of the solar 
neutrino problem, but for a {\em large mixing angle solution.}  Thus, in this 
tale an essentially correct physical idea, but with some initially incorrect 
specific assumptions, led eventually to a surprising resolution of a fundamental 
problem. As a philosophical aside, this story represents a beautiful example of 
the scientific method at work.

\section{Additional Teaching Resources}
The material discussed here is a  self-contained introduction to learning and teaching the physics of vacuum and matter solar neutrino oscillations.  However, additional resources are available for those who wish to go further.   The present paper is a synopsis of a more extensive discussion  that may be found in Chapters 10--13 of the book {\em Stars and Stellar Processes} (Mike Guidry, Cambridge University Press, 2019).  Those chapters contain  25 problems relevant to the present material that go into more technical depth than there is room for here, with complete solutions available online to instructors, and a subset of solutions available online to students,   Color lecture slides also are available to instructors from the publisher that are suitable for teaching the present material.

\section{Summary}

A concise introduction  to neutrino vacuum and matter oscillations in a two-flavor 
model has been presented.  Sample calculations with this formalism, in 
concert with observational data, demonstrate explicitly the 
resolution of the solar neutrino problem through neutrino flavor oscillations 
and the associated MSW matter resonance.  The intent of this presentation has been to make available to  instructors in classes such as astrophysics and quantum mechanics at the advanced undergraduate and beginning graduate level, and to motivated students through self-study, an introduction to the theory of solar neutrino oscillations that assumes only a basic knowledge of quantum mechanics, calculus, differential equations, and  $2\times 2$ matrices, and assumes no special prior knowledge of elementary particle physics, quantum field theory, or solar astrophysics.

\appendix*   % Omit the * if there's more than one appendix.

\section{Natural Units}

The formalism described here uses {\em natural units,} 
chosen such that $\hbar=c=1$. Such units are particularly convenient for 
relativistic quantum field theories and are ubiquitous in the literature  of neutrino oscillations.  Conversion between natural units and more 
standard ``engineering units'' is basically a dimensional analysis problem.  
Let's illustrate with an example.
Consider the neutrino oscillation length $L$.  Multiply the first expression in \eq{vosc8} (which is $L$ expressed in $\hbar = c = 1$ units) by 
$c^4/c^4 = 1$ to give
$$
L = \frac{4\pi E c^4}{\Delta m^2 c^4} = \frac{4\pi E c^4}{\Delta E^2} ,
$$
where $\Delta E^2 \equiv \Delta m^2 c^4$ has units of energy squared.  Let 
$[x]$ denote the units of a variable $x$ and define our standard length unit as $\Script 
L$, our standard energy unit as $\Script E$, and our standard time unit as 
$\Script T$. Then dimensionally,
$$
[L] %= %\left[ \frac{4\pi E c^4}{\Delta E^2} \right]
= \frac{ [E][c^4]}{[\Delta E^2]} 
= \frac{\Script E \Script L^4 \Script T^{-4}}{\Script E^2}
= \Script L \frac{\Script L^3}{\Script E \Script T^4}.
$$
The final expression should have units of length $\Script L$ in normal units, so the preceding 
result in natural units must be multiplied by a combination of $\hbar$ and $c$ having units of 
$\Script E \Script T^4/ \Script L^3$ to convert to normal units.  Since in normal units
$$
[\hbar] = \Script E \Script T 
\qquad [c] = \Script L/\Script T
$$
the required factor is $\hbar/c^3$ and 
\begin{align*}
   L &= \frac{\hbar}{c^3} \times \frac{4\pi E}{\Delta m^2}
= \frac{4\pi E \hbar c}{\Delta m^2 c^4}
\\
     &= 2.48 \times 10^{-3} \left( \frac{E}{\units{MeV}}\right)
      \left( \frac{\units{eV}^2}{\Delta E^2}\right) \units{km},
\end{align*}
where the neutrino energy $E$ is in MeV and the energy squared difference 
$\Delta E^2$ corresponding to the  mass squared difference $\Delta m^2$ is in 
eV$^2$.
Transformation between natural units and everyday (engineering) units for other 
quantities may be carried out in a similar way.

\begin{acknowledgments}

Discussions of this material with Bill Bugg and Eirik Endeve,  
and 
with the students in Astronomy 411 and Astrophysics 615 classes at the 
University of Tennessee are 
gratefully acknowleged. This work was partially supported by LightCone 
Interactive LLC.

This work has been supported by the US Department of Energy and the Oak Ridge National Laboratory. ORNL is managed by UT-Battelle, LLC, for the US Department of Energy under contract no. DE-AC05-00OR22725.

\end{acknowledgments}

\end{document}